\documentclass{aa}  

\usepackage{graphicx}
\usepackage{txfonts}
\usepackage[colorlinks,citecolor=blue,]{hyperref}

\usepackage{amsmath}	
\usepackage{verbatim,xspace,pifont,array,multirow,bigints}
\usepackage{subfigure}
\usepackage[utf8]{inputenc}

\usepackage[dvipsnames]{xcolor}

\newcommand{\gr}{$\gamma$-ray\xspace}
\newcommand{\grs}{$\gamma$\,rays\xspace}
\newcommand{\g}{$\gamma$\xspace}
\newcommand{\pg}{p$\gamma$\xspace}
\newcommand{\integral}{INTEGRAL\xspace}
\newcommand{\spi}{\integral/SPI\xspace}
\newcommand{\hess}{H.E.S.S.\xspace}
\newcommand{\blj}{\texttt{BHJet}\xspace}
\newcommand{\mlj}{\texttt{MLJet}\xspace}
\newcommand{\fermi}{\textit{Fermi}/LAT\xspace}

\newcommand{\bh}{BHXB\xspace}
\newcommand{\bhs}{BHXBs\xspace}
\newcommand{\src}{A0620--00\xspace}
\newcommand{\dragon}{\texttt{DRAGON2}\xspace}

\defcitealias{kantzas2020cyg}{K21}

\defcitealias{kantzas2022gx}{K22}
\newcommand{\kgx}{{\citetalias{kantzas2022gx}}\xspace}

\begin{document}

   \title{Quiescent black hole X-ray binaries as multi-messenger sources}
   \titlerunning{q\bhs as multi-messenger sources}

   \author{Dimitrios Kantzas
          \and
          Francesca Calore
          }
    \authorrunning{Kantzas \& Calore}
   \institute{Laboratoire d’Annecy-le-Vieux de Physique Th\'eorique (LAPTh), USMB, CNRS, 74940 Annecy, France\\
              \email{dimitrios.kantzas@lapth.cnrs.fr\\}
             }

   \date{Received --; accepted --}

 
  \abstract
   { The origin of Galactic cosmic rays (CRs) is unknown even though they have traditionally been connected to supernovae based on energetic arguments. In the past decades, Galactic black holes in X-ray binaries (BHXBs) have been proposed as candidate sources of CRs, which revises the CR paradigm.
    \bhs launch two relativistic jets during their outbursts, but recent observations suggested that these jets may be launched even during quiescence. \src is a well-studied object that shows indications of jet emission. We study the simultaneous radio-to-X-ray spectrum of this source that was detected while the source was in quiescence to better constrain the jet dynamics. Because most \bhs spend their lifetimes in quiescence (q\bhs), we used the jet dynamics of \src to study a population of $10^5$ such sources distributed throughout the Galactic disc, and a further $10^4$ sources that are located in the boxy bulge around the Galactic centre. 
    While the contribution to the CR spectrum is suppressed, we find that the cumulative intrinsic emission of q\bhs from both the boxy bulge and from the Galactic disc adds to the diffuse emission that various facilities detected from radio to TeV \grs. We examined the contribution of q\bhs to the Galactic diffuse emission and investigated the possibility of SKA, \integral, and CTAO to detect individual sources in the future. 
    Finally, we compare the predicted neutrino flux to the recently presented Galactic diffuse neutrino emission by IceCube.
}

   \maketitle
%

\section{Introduction}

Cosmic rays (CRs) are charged particles with an extraterrestrial origin. Despite decades of research, we still lack a full understanding of the CR sources and the physical mechanism behind their acceleration. 
When CRs accelerate, they reach high energies that allow the emission of \grs. On the one hand, leptonic CRs can upscatter background radiation to \grs, and on the other hand, hadronic CRs can interact inelastically with background photons and/or gas to produce secondary particles, such as \grs and neutrinos \citep{mannheim1994interactions}. 
When Galactic CRs propagate in the interstellar medium, they contribute to the diffuse emission that is detected along the Galactic plane from soft X-rays to very-high energy TeV \grs through similar processes
\citep{perez2019galactic,Ackermann_2012,LHAASO2023Galactic,Amenomori2021first}.

The X-ray background especially in the 1-100\,keV energy range is relatively well studied, for instance. The bulk of this emission, about 90\%, is thought to mainly originate in cataclysmic events, while the remaining 10\% can be produced by unresolved sources, which indicates CR acceleration \citep{perez2019galactic}. 
Likewise, for the diffuse background of more energetic MeV X-rays, some unresolved Galactic sources are required to explain the entire spectrum \citep{berteaud2022strong}. Recently, pulsars were suggested as promising MeV emitters, but due to a lack of a radio counterpart, q\bhs cannot be ruled out so far. At higher energies, in the GeV and TeV bands, the picture is similarly unclear. In particular, the diffuse GeV background \citep{Ackermann_2012}, especially towards the Galactic bulge, may also be caused by dark matter annihilation and/or decay, which complicates the search for the CR accelerators even further \citep{Dodelson2008Extracting,Calore2015tale}. 
Similarly, Galactic sources that emit at TeV energies and beyond are among the most powerful Galactic CR accelerators. They probably cause the diffuse emission that was recently measured by the Large High Altitude Air Shower Observatory (LHAASO; \citealt{LHAASO2023Galactic}) and Tibet Air Shower-\g (AS-$\gamma$; \citealt{Amenomori2021first}). A final evidence for Galactic CR accelerators comes from the Galactic diffuse neutrino emission. This evidence remains preliminary so far, but indicates specific types of sources \citep{icecube2023observation}.

Traditionally, the explosive deaths of massive stars have been considered to be Galactic CR sources \citep{baade1934cosmic, Ginzburg1964,Blasi2013}. Numerous works have recently suggested stellar mass black holes (BHs) in X-ray binaries (XRBs; \bhs, henceforth) as promising candidates for CR sources \citep{romero2003hadronic,fender2005CRXRBs,BEDNAREK2005galactic,Torres_2005,reynoso2008ss433,cooper2020xrbcrs,carulli2021neutrino}. 
In particular, stellar-mass BHs in XRBs accrete mass from the companion star, and when they go into outburst, they launch two relativistic jets \citep{mirabel1994superluminal,corbel2000coupling,Corbel2013universal,fender2001powerful,Corbel2002NIR,fender2004unified,Fender2009jets,mcclintock2006compact}. These jets can accelerate particles to high energies, which leave their imprint in the radio-to-\gr regime, such as the cases of Cygnus~X--1 \citep{gallo2005dark,zanin2016detection}, Cygnus~X--3 \citep{miller2004time,Tavani_2009CygnusX3}, and more recently, SS433, which was detected even in the PeV \citep{abeysekara2018very,Safi-Harb_2022}.

The \bhs are observed to launch jets during outbursts, mainly during the so-called hard X-ray spectral state \citep{mcclintock2006compact}, but sometimes also during the soft state (see e.g. the case of Cygnus~X--3; \citealt{koljonen2010hardness,Cangemi2021integral}). Recent simultaneous radio to X-ray observations showed evidence of a flat radio spectrum along with a hard X-ray spectrum from several \bhs in quiescence (q\bhs; \citealt{Gallo_2007Quiescent,maitra2009constraining,plotkin2013x,Plotkin2014Constraints,Plotkin_2017TRANSITION,Gallo2014lowest,connors2016mass,dinccer2017multiwavength,dePolo2022flickering}). These spectral features indicate the existence of a pair of relativistic jets that carry significantly less power than the jets during the outburst, however.

To better capture the jet physics and understand the particle acceleration along the jets, several models were suggested in the past \citep{tavecchio1998constraints,mastichiadis_kirk_2002,marscher2008inner,Tramacere2009jetset}. Based on the nominal work of \cite{blandford1979relativistic}, a multi-zone jet prescription can reproduce the multi-wavelength spectral emission we detect from these sources, and it can describe the jet morphology that numerous sources demonstrate \citep[see e.g.][]{Janssen2021ehtCenA}. 
In this work, we use the multi-zone jet model as initially sketched by \cite{markoff2001jet} and \cite{Markoff2005} and further developed by \cite{lucchini2022BHJet} for a purely leptonic non-thermal emission and by \cite{kantzas2020cyg}, who expanded it to lepto-hadronic radiative processes. More precisely, we use and compare two different jet scenarios. In the first scenario, we expand the jet dynamics of \cite{lucchini2022BHJet} to include the hadronic interactions as described in \cite{kantzas2020cyg}, and we refer to this model as \blj. The second scenario is the mass-loading case of \cite{Kantzas2023MassLoading}, which was developed to address the proton-power problem, that is, the question of the origin of the energy excess that is used by protons to accelerate to high energies \citep{boettcher2013leptohadronic,Liodakis2020}, and we refer to this model as \mlj. 

Even thought they are faint and many times fail to exceed the observability threshold, a few q\bhs are still detected. \src, for instance, demonstrates a flat radio spectrum, and recent modelling of the X-ray spectrum did not rule out some significant jet contribution \citep{Gallo_2007Quiescent,connors2016mass,dinccer2017multiwavength,dePolo2022flickering}. This broadband coverage allows us to develop more robust constraints of the jet contribution to the electromagnetic spectrum to better understand their dynamical properties. Based on the findings of the multi-wavelength spectral fitting of \src, we investigated the possible contribution of the entire population of q\bhs to the CR spectrum detected on Earth because some dozens to hundreds of thousands of these sources may reside in the Milky Way (see e.g. \citealt{Olejak2019synthesis} and references below). 
We show below that the contribution to the Galactic CR spectrum is negligible, but a further, indirect, indication for CR acceleration is the total contribution of q\bh jets to the observed diffuse emission in multiple frequencies.

In Section~\ref{Sec: Jet model}, we describe the jet model we used in this work, and we apply it to \src in Section~\ref{Sec: the prototypical case}. We investigate the possible contribution of the entire population of q\bhs to the diffuse emission in Section~\ref{Sec: the intrinsic emission from all sources}, and finally, we discuss the results of this work in Section~\ref{Sec: discussion}, and we conclude in Section~\ref{Sec: summary and conclusions}.

\section{Jet emission}\label{Sec: Jet model}

\subsection{\blj}
The jets launched by \bhs have similar physical properties as
active galactic nuclei (AGN) jets, but on much smaller scales \citep{Merloni2003fundamental,Falcke2004FP}. In this work, we further develop the multi-zone jet model that describes the flat to inverted radio spectra observed for \bhs \citep[see e.g.][]{russell2014multiwavelength} that was initially presented by \cite{falcke1994jet} and \cite{markoff2001jet,markoff2003exploring,Markoff2005}. The most recent and user-friendly version of this model, referred to as \blj, is thoroughly described in \cite{lucchini2022BHJet} and can be found in an online repository\footnote{\href{https://github.com/matteolucchini1/BHJet}{https://github.com/matteolucchini1/BHJet}}. 

\blj solves the transport equation of a mixed population of thermal and non-thermal electrons along the jet axis and estimates synchrotron and inverse-Compton scattering (IC) emissions. This complex but still physically motivated jet prescription can prove useful in understanding the jet kinematics of not only small-scale jets (see e.g. \citealt{maitra2009constraining,Plotkin2011fundamental,Plotkin2014Constraints,Markoff_2015,Connors2019combining,Lucchini2021correlation,kantzas2020cyg,kantzas2022gx}), but also large-scale AGN jets (see e.g. \citealt{lucchini2019breaking,lucchini2019unique}). In either case, two fundamental aspects can explain the electromagnetic constraints well. Firstly, a Poynting-flux-dominated thermal but still relativistic jet base explains the infrared (IR) observations well, especially those that indicate a jet contribution (see e.g. \citealt{Gallo_2007Quiescent,Gandhi2011synchbreak} and references above). This jet base is described by its initial radius $r_0$, the injected power $P_{\rm jet}$, the temperature of the relativistic electrons $T_e$, and the equipartition arguments. $r_0$, $P_{\rm jet}$, and $T_e$ are free parameters, and for simplicity, we assumed that the plasma $\beta$, defined as the energy density of the gas over the energy density of the magnetic field, is equal to 0.02 (see \citealt{lucchini2022BHJet} for a detailed explanation). Finally, the exact composition of the jet is not fully understood, but the simple assumption of an equal number of electrons and protons is broadly accepted. 

Far beyond the jet base and at some distance that usually reaches dozens to thousands of gravitational radii \footnote{one gravitational radius is defined as $r_g  = GM_{\rm bh}/c^2 \simeq 1.5\times 10^5\,(M_{\rm bh}/M_{\odot})\,\rm cm $, where $G$ is the gravitational constant, $M_{\rm bh}$ is the mass of the black hole, $c$ is the speed of light, and $M_{\odot}$ is the solar mass.}, a fraction of the thermal electrons start to accelerate to a non-thermal power law in energy due to some particle acceleration mechanism that is unknown so far. This region is called particle acceleration region and is located at some distance $z_{\rm diss}$ . The jet energy is here further dissipated into the bulk velocity, the magnetisation (defined as $\sigma = B^2/4\pi\rho c^2$, where $B$ is the magnetic field strength and $\rho$ mass density of the jet segment), and the particles. This location is also the transition between an optically thin to a thick jet plasma that explains the spectral break that is usually detected in the IR band of the spectrum (see e.g. \citealt{Gandhi2011synchbreak,Russell2014J1836}). Because the particle acceleration region is tightly connected to the jet base, its dynamical properties such as the radius and the strength of the magnetic field depend on its distance from the BH, which is a free parameter.  

As already mentioned, we do not know the particle acceleration mechanism, which can significantly alter the observational imprints, however. To better investigate the effect of this particle acceleration, but avoid increasing the number of free parameters, we assumed that the minimum electron energy of the power law is the peak of the thermal distribution at $T_e$, and the index $p$ remains a free parameter. We calculated the maximum energy of the non-thermal electrons by equating the characteristic timescales of the acceleration and the radiative losses. For an efficient particle acceleration and the case of \bhs, the non-thermal electrons reach energies of some dozen GeV. These energetic electrons in the strong magnetic fields of the jets can produce a hard synchrotron spectrum that shines up to X-rays. Further upscattering of these X-rays by the non-thermal electrons can explain the GeV \grs that are detected from these sources \citep{Tavani_2009CygnusX3,zanin2016detection}.

A purely leptonic jet model such as the one we discussed so far can explain the overall electromagnetic emission detected by \bh and AGN jets sufficiently, but it cannot treat the acceleration of hadronic particles that is likely to occur inside these sources because of the correlation to astrophysical neutrinos \citep{Keivani_2018}. To better investigate the hadronic acceleration and the contribution of flaring \bhs to the CR spectrum, we developed a lepto-hadronic jet model in \cite{kantzas2020cyg}. The jet dynamics assumed in that work was a pressure-driven jet that may even be dominated by particles because the particle energy density dominates the Poynting flux. For the first time, to allow for a Poynting flux-dominated jet that includes the inelastic hadronic processes, we combined the jet dynamics of \blj with the hadronic processes discussed in \cite{kantzas2020cyg}. More specifically, similar to the non-thermal electrons, we assumed that the protons are initially cold. In the dissipation region, a fraction of protons populates a power law in energy from some minimum energy of 1\, GeV up to some maximum energy that was self-consistently calculated per jet segment, following a similar prescription as the Hillas criterion \citep{hillas1984origin,jokipii1987rate}. The power-law index remained a free parameter that was the same for electrons and protons for simplicity. The non-thermal protons of each jet segment interact inelastically with the cold protons of the jet flow, the radiation of this specific jet segment, and other photon and gas fields such as the companion star radiation field and its stellar wind. The proton-proton (pp) and photohadronic (\pg) interactions lead to the formation of distributions of charged and neutral pions that eventually decay into \grs, secondary electrons, and neutrinos \citep{mannheim1993proton,mannheim1994interactions,rachen1993extragalactic,rachen1998photohadronic,MUCKE2003protonBLLac}. These interactions are complex and therefore require Monte Carlo simulations to properly produce the secondary populations. Time- and source-consuming processes like this would not allow for a fast comparison with observational data, however. We therefore chose to use the semi-analytical formulae of \cite{kelner2006energy} for pp and that of \cite{kelner2008energy} for \pg, respectively.

\subsection{\mlj}

Recent numerical magnetohydrodynamic simulations in the general relativity regime (GRMHD) show that not only a Poynting-flux-dominated jet is a natural output of the \cite{blandford1977extraction} and \cite{Blandford1982hydromagnetic} launching mechanisms, but that a significant fraction of the energy is also dissipated to particles. For instance, \cite{chatterjee2019accelerating} performed one of the highest 2D resolution GRMHD simulations to show that instabilities in the interface between the jet edge and the wind of the accretion disc allow for eddies that transport matter from the wind to the jet. Hybrid GRMHD and particle-in-cell simulations of these setups proved that particle acceleration indeed occurs and that particles gain non-thermal energies \citep{Sironi2020}. 

\cite{Kantzas2023MassLoading} captured the macroscopic picture of this mass-loading scenario. We adopted the jet dynamics from the GRMHD simulations and combined it with the lepto-hadronic processes. We refer to this scenario as \mlj.
More precisely, we parametrised the profiles along the jet axis of the bulk velocity, the magnetisation, and the specific enthalpy, which is an estimate of the excess energy that is converted into non-thermal protons. The initial setup was again driven by the physics of the jet base and the jet region, where the mass-loading starts to become strong (we adopted the same parameter as \blj, namely $z_{\rm diss}$). The mass-loading was initiated at a distance $z_{\rm diss}$ from the BH and stopped at $\sim 10\,z_{\rm diss}$ (see the discussion in \citealt{chatterjee2019accelerating} and \cite{Kantzas2023MassLoading}). 
A further vital aspect is the initial magnetisation at the jet base because this amount of energy does not only lead to the bulk jet acceleration, but also allows for more abundant hadronic counterparts. We adopted a relatively medium value of $\sigma = 10$ and a bulk Lorentz factor after reaching a maximum value of 3 at $z_{\rm diss}$. Finally, a further important macroscopic quantity is the jet composition. In \mlj, we assumed that the jet was launched lepton-dominated, and the ratio of electrons to protons was a free parameter $\eta = n_e/n_p$, where  $n_e$ is the number density of pairs of electrons, and $n_p$ is the number density of protons. At the highest mass-loading, we assumed an equal number of electrons and protons to entrain the jets, modifying their dynamics. This assumption ensured that the jet remained charge-neutral despite the mass-loading. All the free parameters used in this work, and in particular, for the case of \src, are tabulated in Table~\ref{table: free parameters}.

\section{Studying the prototypical case of \src}\label{Sec: the prototypical case}

\begin{figure}
    \centering
    \includegraphics[width=1\columnwidth]{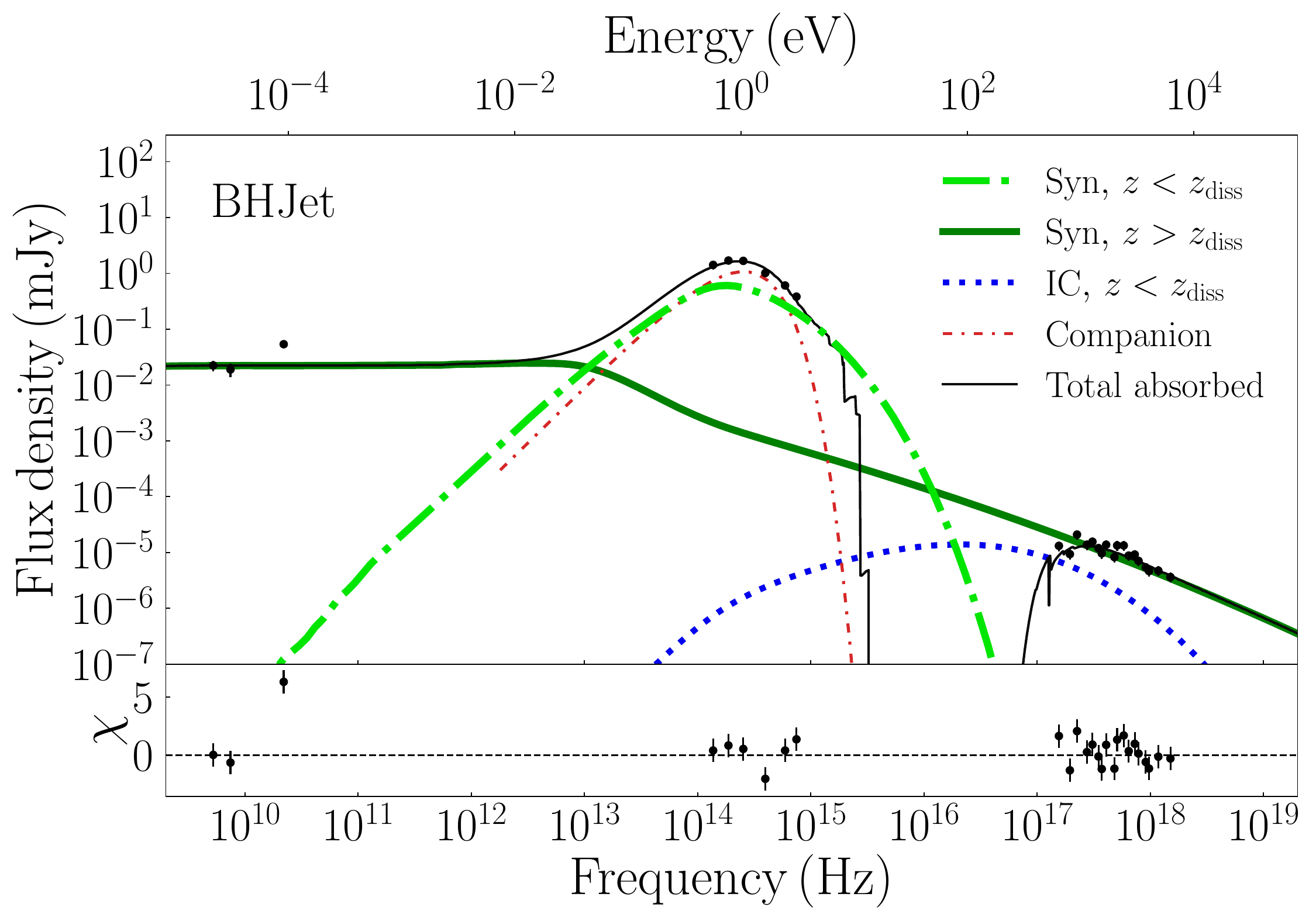}\\
    \includegraphics[width=1\columnwidth]{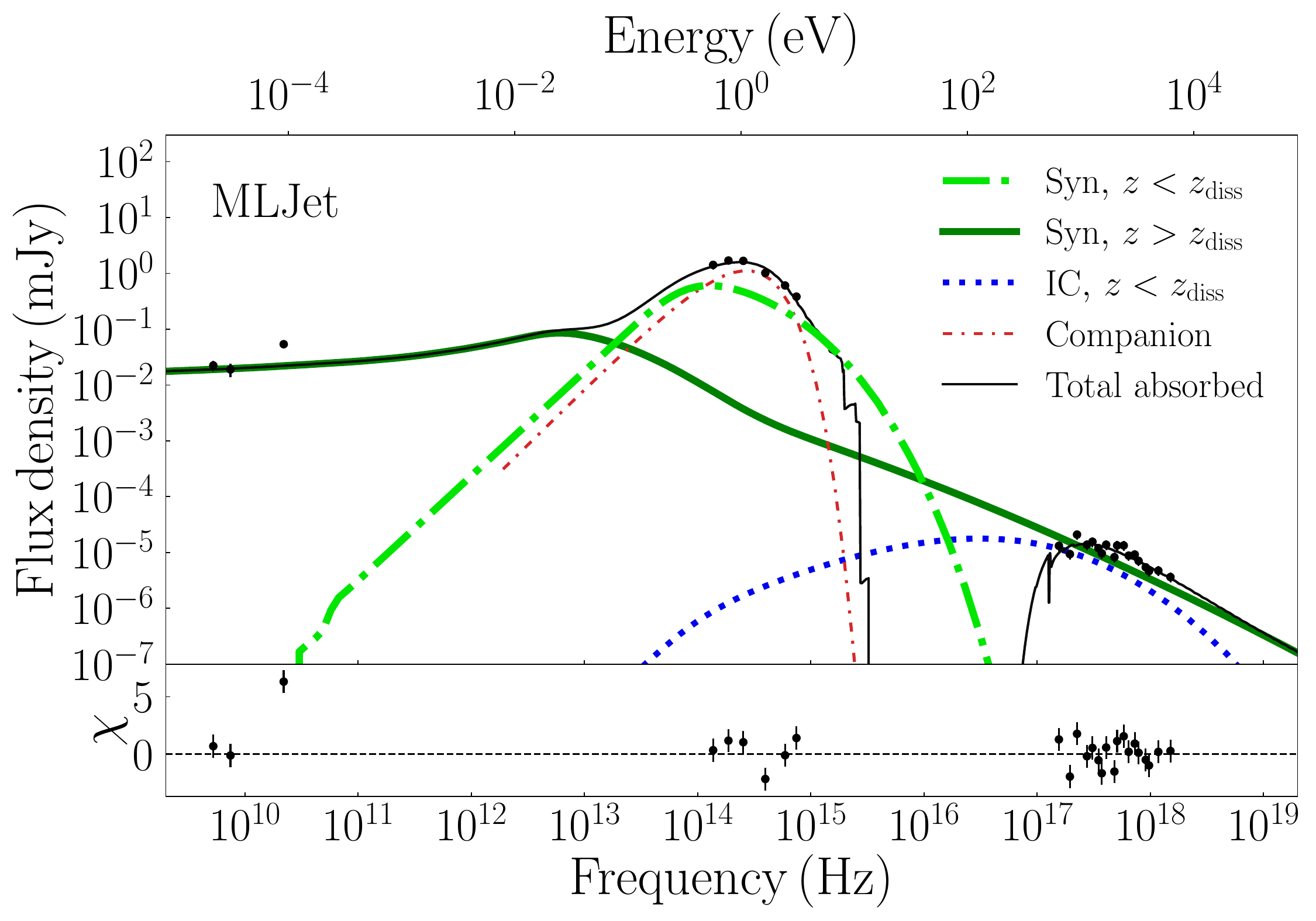}
    \caption{Best fit with residuals of the multi-wavelength flux density of the 2013 observations of \src from \protect\cite{dinccer2017multiwavength}. In the upper panel, we show the result for the case of \blj, and in the lower panel, we show the case of \mlj. The solid black line shows the total absorbed emission, and the individual components are explained in the legend.}
    \label{fig: leptonic flux density synchrotron dominated X rays}
\end{figure}

\src is a typical q\bh located at a distance of $1.06\pm 0.12$\,kpc \citep{Cantrell_2010}. The mass of the BH is estimated to be $6.61\pm 0.25\,\rm M_{\odot}$ and the inclination of the system is $51\pm 1 ^{\circ}$ \citep{Cantrell_2010}. We used the multi-wavelength observations of \cite{dinccer2017multiwavength}, which cover the entire electromagnetic spectrum from radio (VLA) to optical/near-infrared (SMARTS telescope) and X-rays with \textit{Chandra}. We used the observation performed on 2013 December 9, and we plot the output in Fig.~\ref{fig: leptonic flux density synchrotron dominated X rays}. In this figure, in particular, we plot the best fit of \blj to the observational data in the upper panel and the best fit of \mlj in the lower panel. To determine the best fit for either jet model, we used the interactive spectral interpretation system (\texttt{ISIS}; \citealt{houck2000ISIS}) to forward-fold the model into X-ray detector space. We used the \texttt{EMCEE} function to explore the parameter space using a Markov chain Monte Carlo method (\citealt{Foreman_Mackey_2013}. 
We performed $10^4$ loops with 20 initial walkers per free parameter, and we rejected the first 50\% of the runs until the method approached a good statistical significance. In Table~\ref{table: free parameters}, we list the free parameters of the two models, along with the results of the best fit and the $1\,\sigma$ uncertainties. In Fig.~\ref{fig: leptonic flux density synchrotron dominated X rays}, we indicate the total flux density to account for absorption, as well as all the individual components from the jet base, the jet, and the companion star, as shown in the legend. The insets in each panel show the residuals of the model. 

\begin{figure}
    \includegraphics[width=1.\columnwidth]{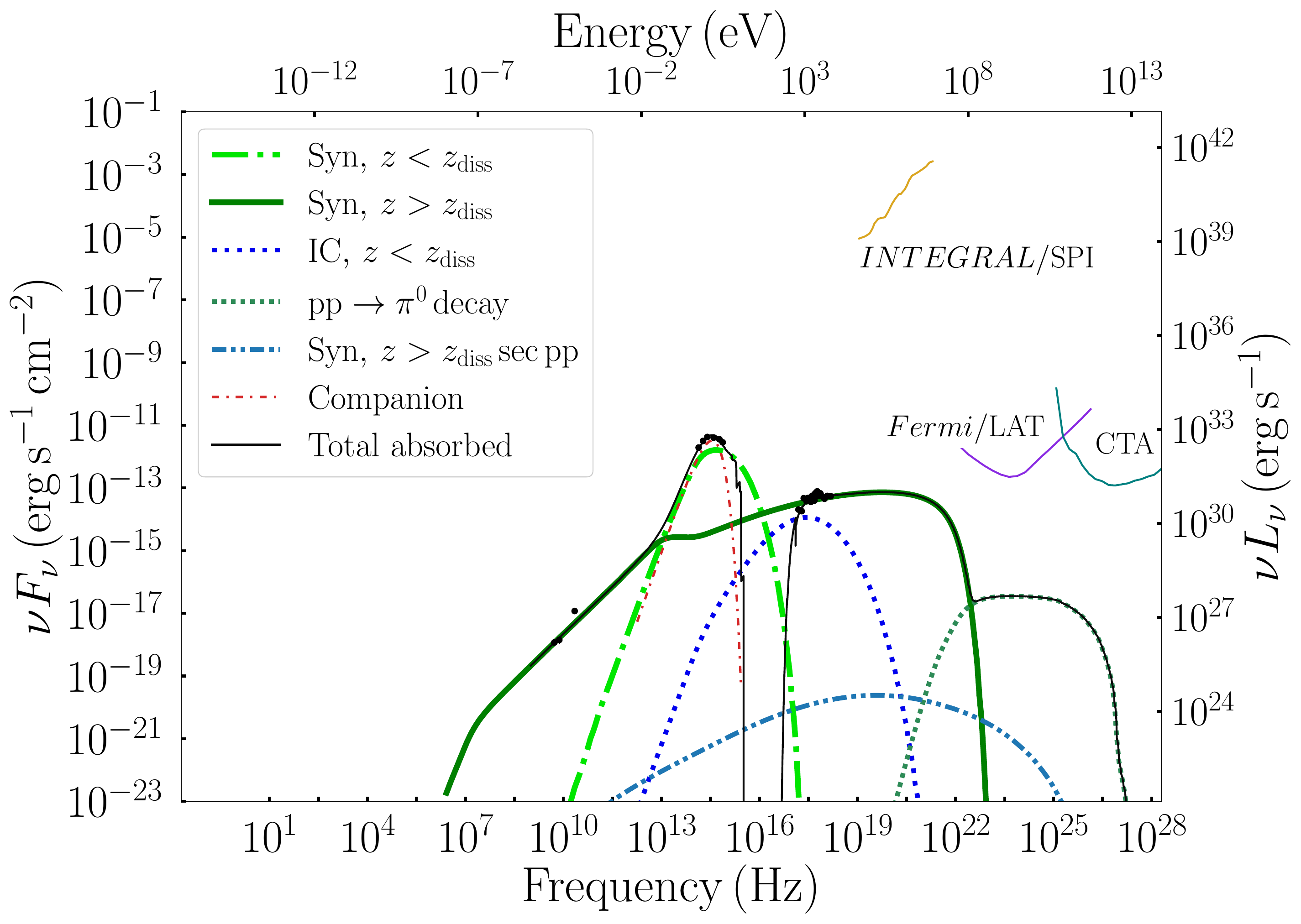}
    \includegraphics[width=1.\columnwidth]{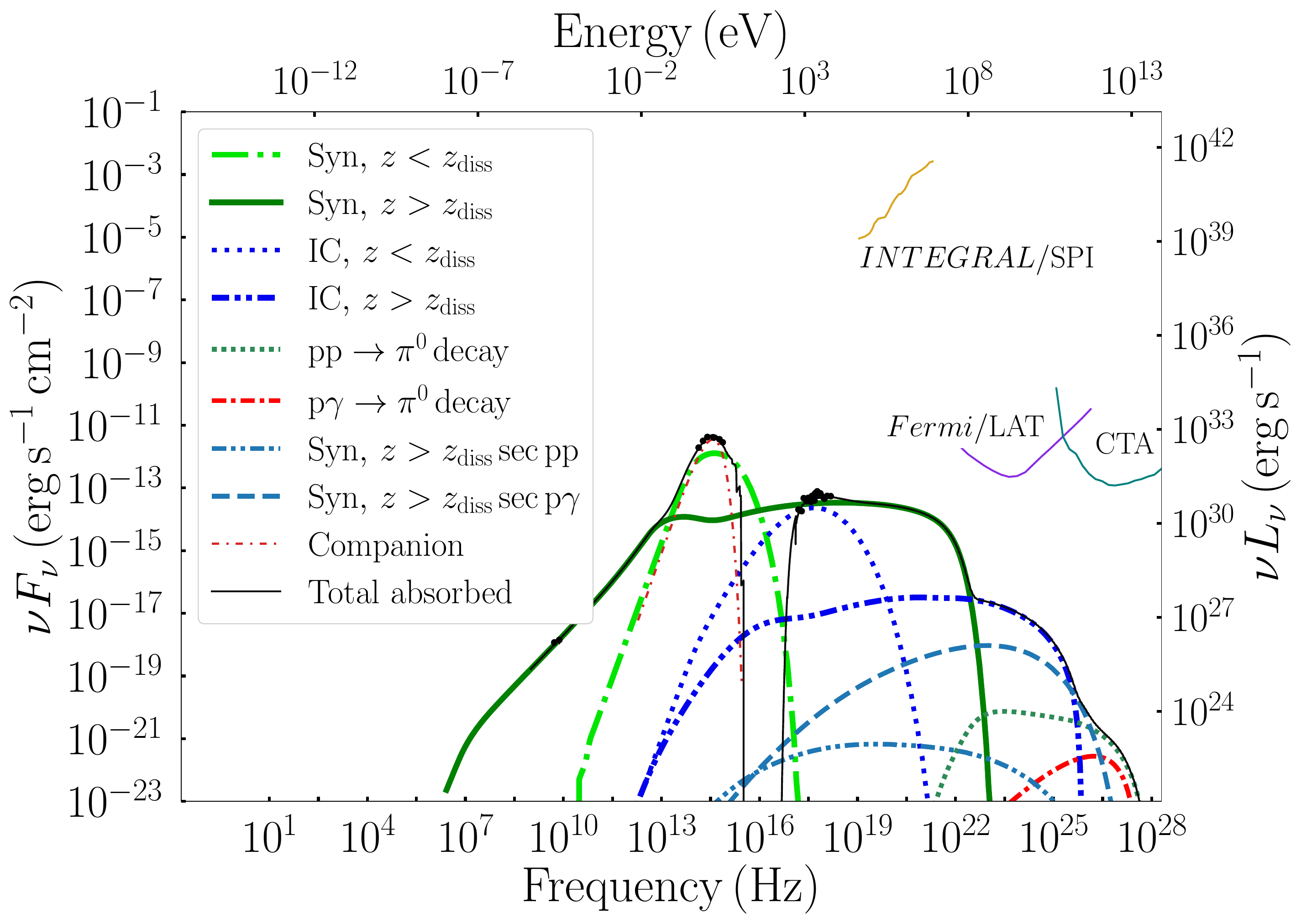}
    \caption{
    Multi-wavelength spectrum of \src for the case of the \blj model in the upper panel and for the case of the \mlj model in the lower panel. The solid black line shows the total emitted spectrum, and the individual components are explained in the legends. The radio-to-X-ray data are the same as in Fig.~\ref{fig: leptonic flux density synchrotron dominated X rays}. We include the \spi, \fermi, and CTAO point source sensitivities for comparison.  }
    \label{fig: hadronic MW SED synchrotron dominated X rays}
\end{figure}

In Fig.~\ref{fig: hadronic MW SED synchrotron dominated X rays}, we show the extended energy spectrum with the radio-to-X-ray data and the sensitivity curves of three instruments in the high-energy regime. We used the sensitivity curve of \spi, the \fermi sensitivity, and the predicted Cherenkov Telescope Array Observatory CTAO sensitivity for the south site. The upper panel  shows the case of \blj, and the lower panel shows the case of \mlj. The contribution from the secondary particles and the individual components is as shown in the legend.

\begin{table*}
\begin{center}
	\setlength{\tabcolsep}{8pt} 
	\renewcommand{\arraystretch}{1.5} 
	\begin{tabular}[b]{lccl}\hline\hline
		\multirow{2}{*}{Parameter} & \multicolumn{2}{c}{Model} & \multirow{2}{*}{Description} 
		\\ & \blj & \mlj &
		
		\\ \hline
        $P_{\rm jet}/\rm L_{Edd}$ & $2.86\pm 0.03\times10^{-5}$ & $7 _{ - 4} ^{+30} \times 10^{-6}$  &Injected power at the jet base\\
        $r_0/r_g$   & $6.3_{-0.5}^{+0.1}$ & $11_{-4}^{+9}$  &Jet-base radius \\
        $z_{\rm diss}/r_g$& $73\pm 1$   & $4 \pm 1$ & Particle acceleration region\\
        $k_BT_e/\rm keV$ & $1800_{-10}^{+40}$ & $2600 _{-100}^{+1700}$ & Jet-base electron temperature \\
        $p$ & $2.070_{-0.001}^{+ 0.003}$ & $2.2 _{-0.1}^{+0.2}$ & Power-law slope of non-thermal particles \\
        $\eta_e$ & 1 (fixed) & $190_{-180}^{+240}$  & pair-to-electron/proton ratio\\
		\hline
	\end{tabular} 
	\caption{The free (fitted) parameters of the multi-wavelength spectral modelling fit to \src. \\
	}\label{table: free parameters}
\end{center}
\end{table*}

\section{Multi-wavelength emission of q\bhs}\label{Sec: the intrinsic emission from all sources}

\subsection{Population model}

\begin{figure*}[!ht]
    \centering
    \subfigure{
        \hspace{-1cm}
        \includegraphics[width=1.\columnwidth]{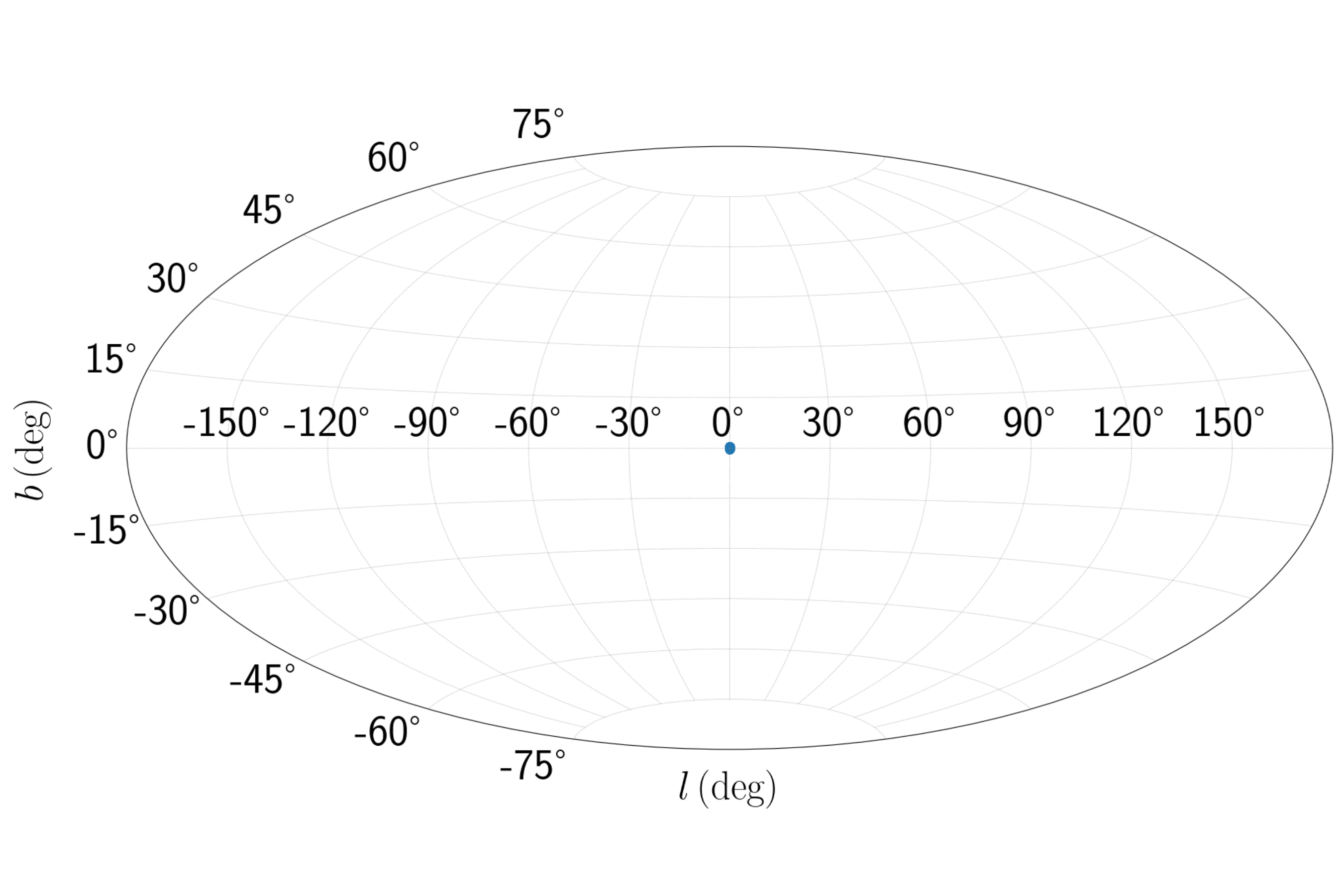}
    }
    \subfigure{
        \includegraphics[width=1.\columnwidth]{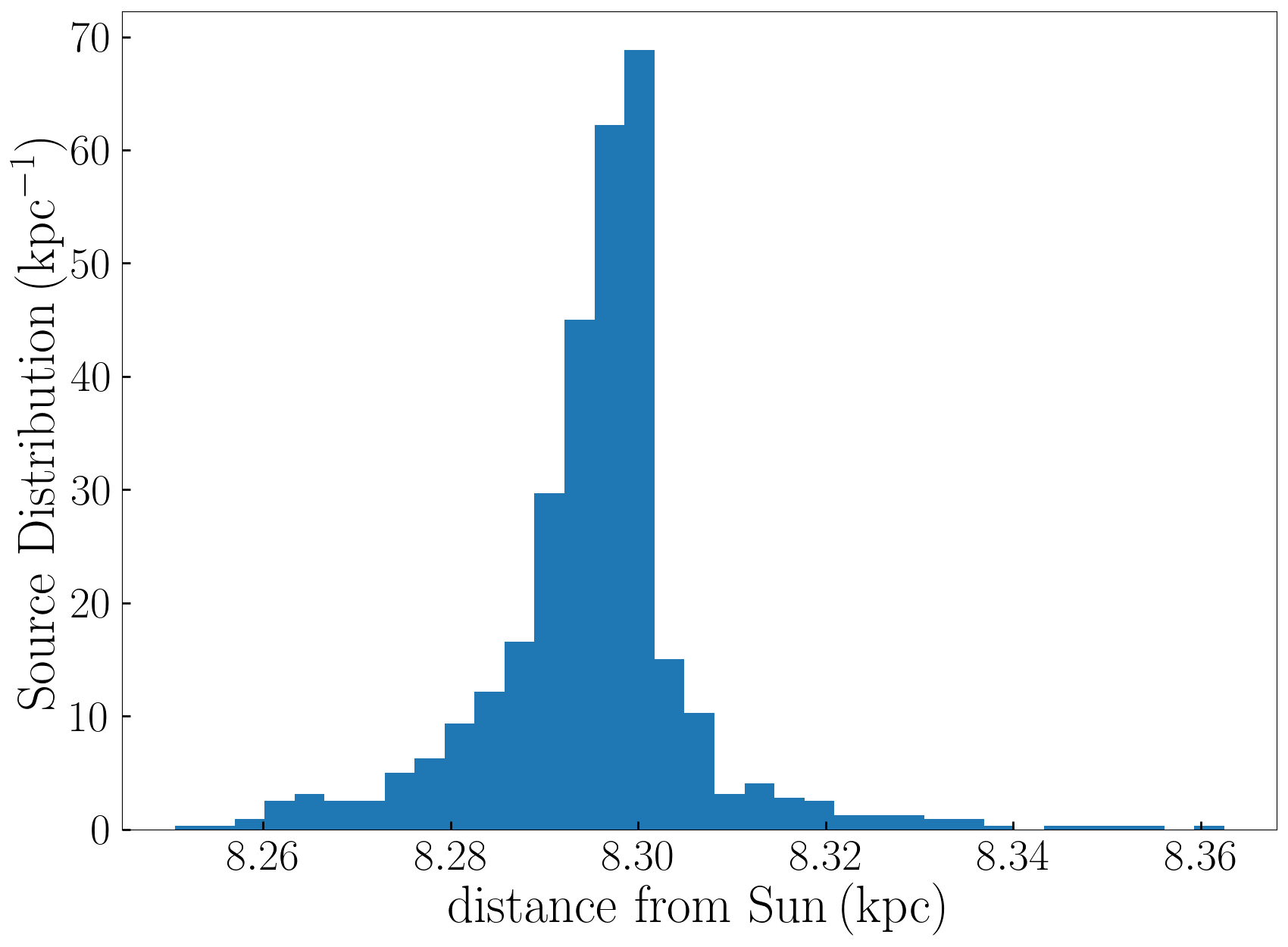}
    }
    \vspace{.25cm}

    \subfigure{
        \hspace{-1cm}
        \includegraphics[width=1.\columnwidth]{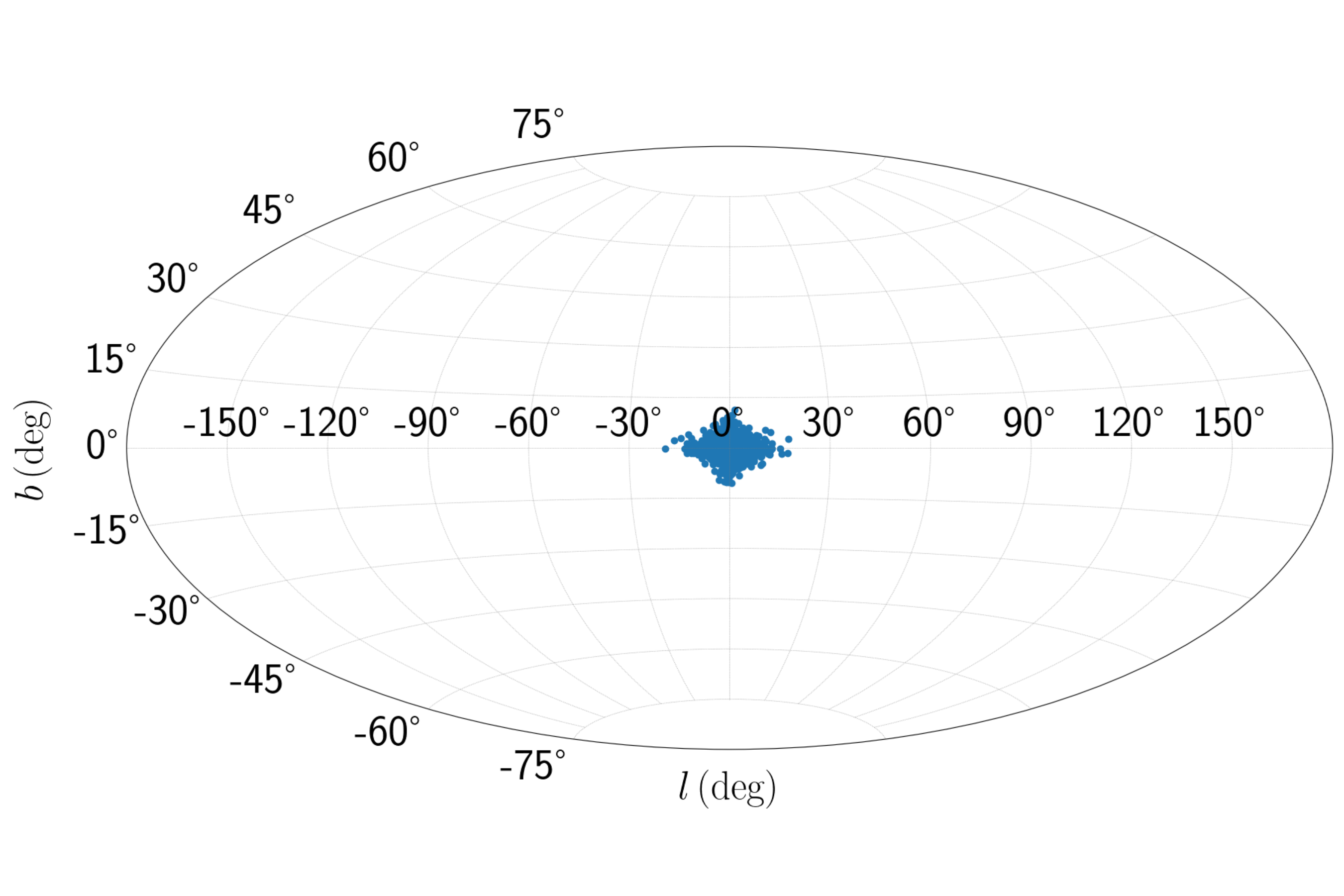}
    }
    \subfigure{
        \includegraphics[width=1.\columnwidth]{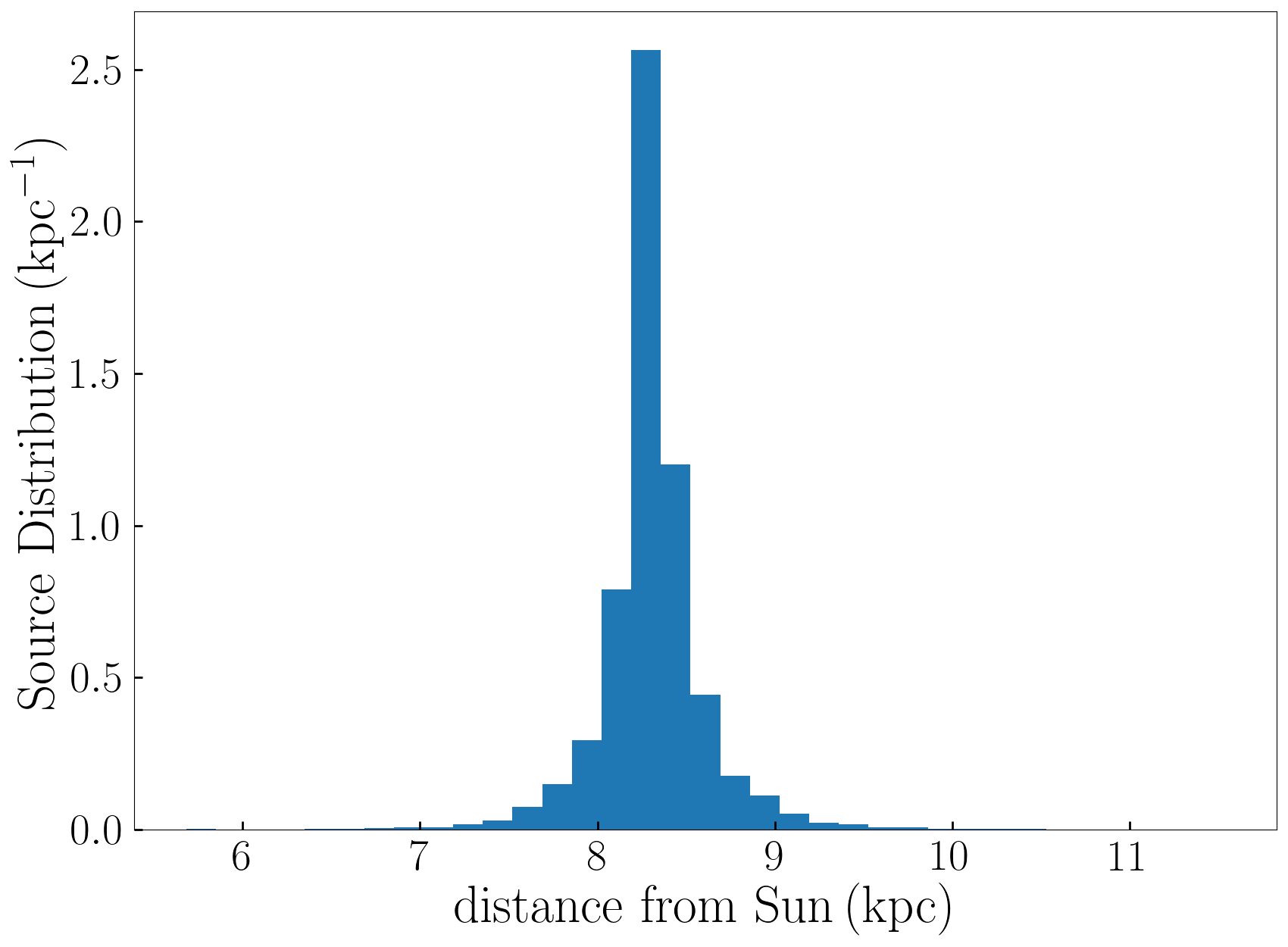}
    }    
    \vspace{-.25cm}
    \subfigure{
        \hspace{-1cm}
        \includegraphics[width=1.\columnwidth]{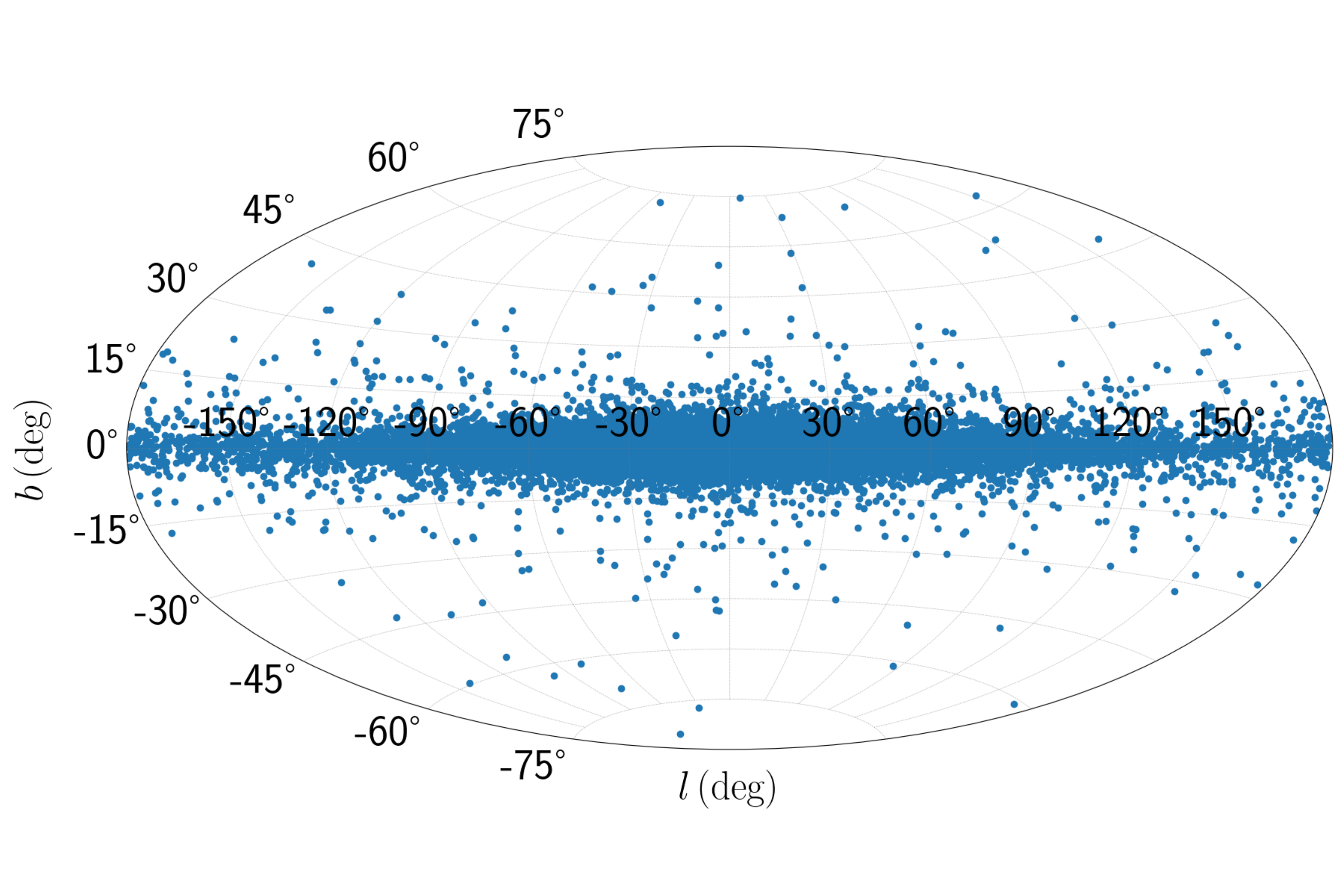}
    }
    \subfigure{
        \includegraphics[width=1.\columnwidth]{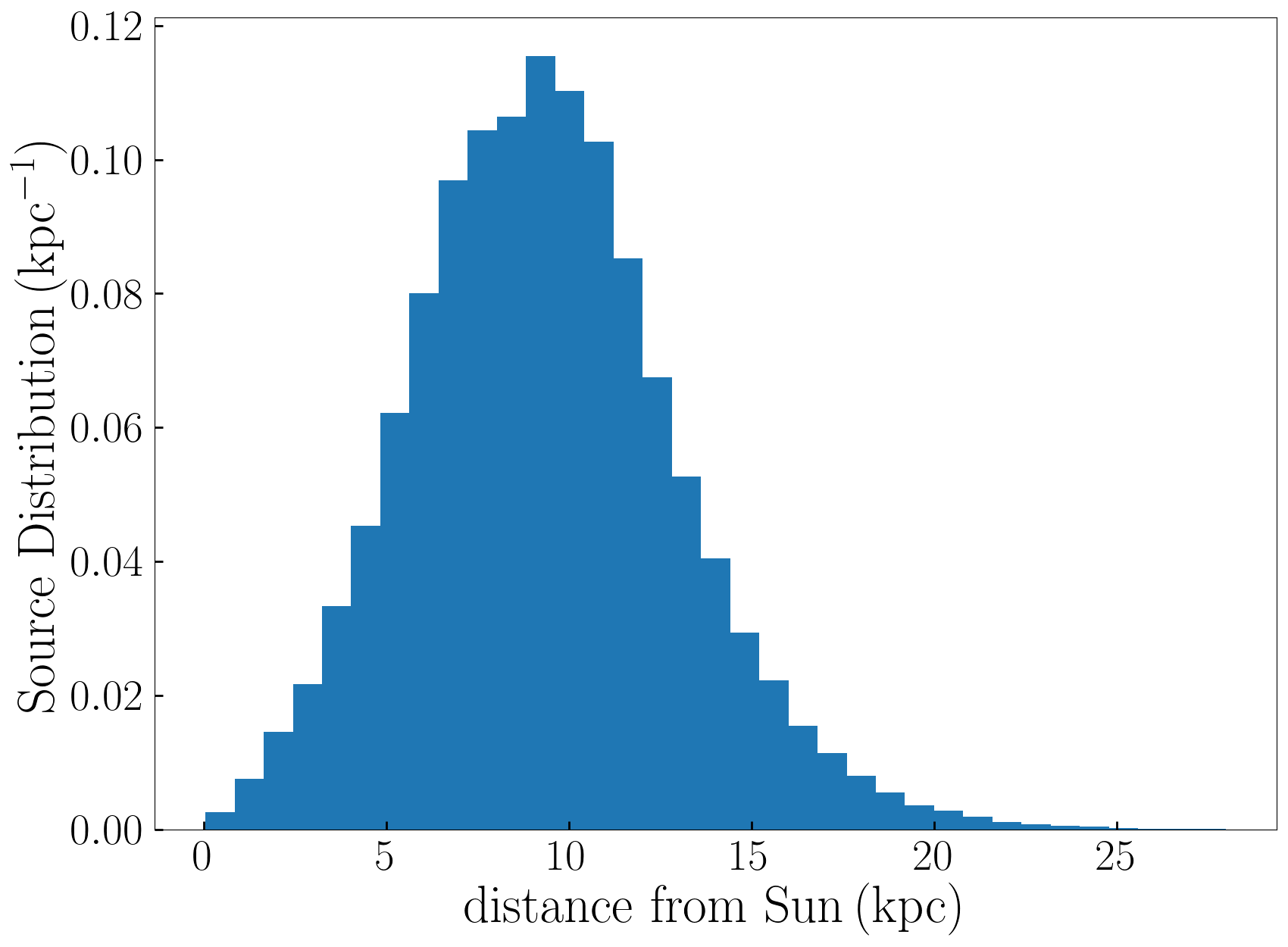}
    }
    \caption{
    Positions of one random realisation of q\bhs in the sky (\textit{left}) in Galactic coordinates and the histogram of their distances (\textit{right}). From top to bottom, we plot the $10^3$ sources at the GC, the $10^4$ boxy bulge, and the $1.2\times 10^4$ Galactic disc.  
 }   
    \label{fig: distances and skymaps}
\end{figure*}

\begin{figure*}[!ht]
    \centering
    \subfigure{
        \includegraphics[width=0.975\columnwidth]{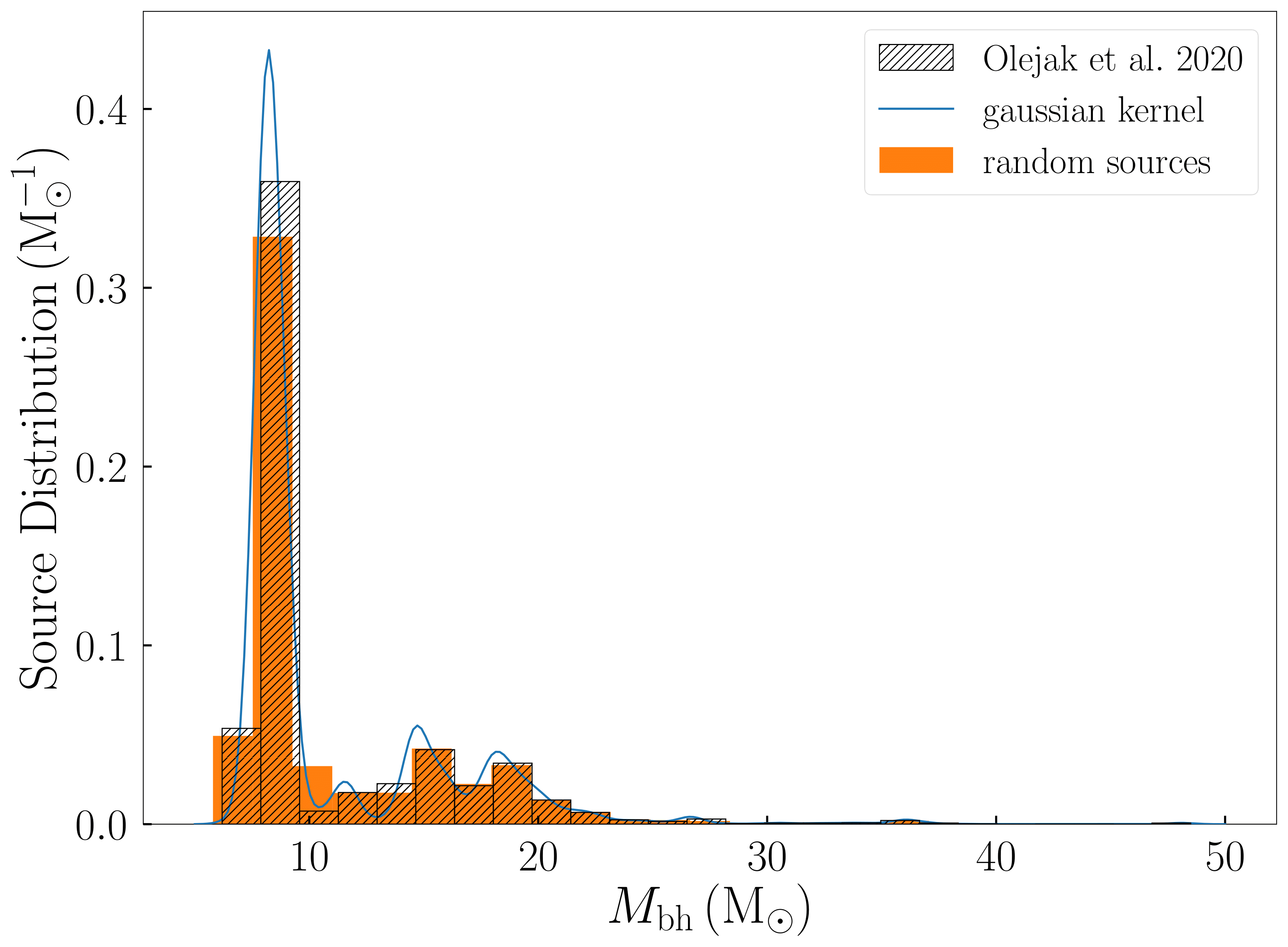}
    }
    \subfigure{
        \includegraphics[width=0.975\columnwidth]{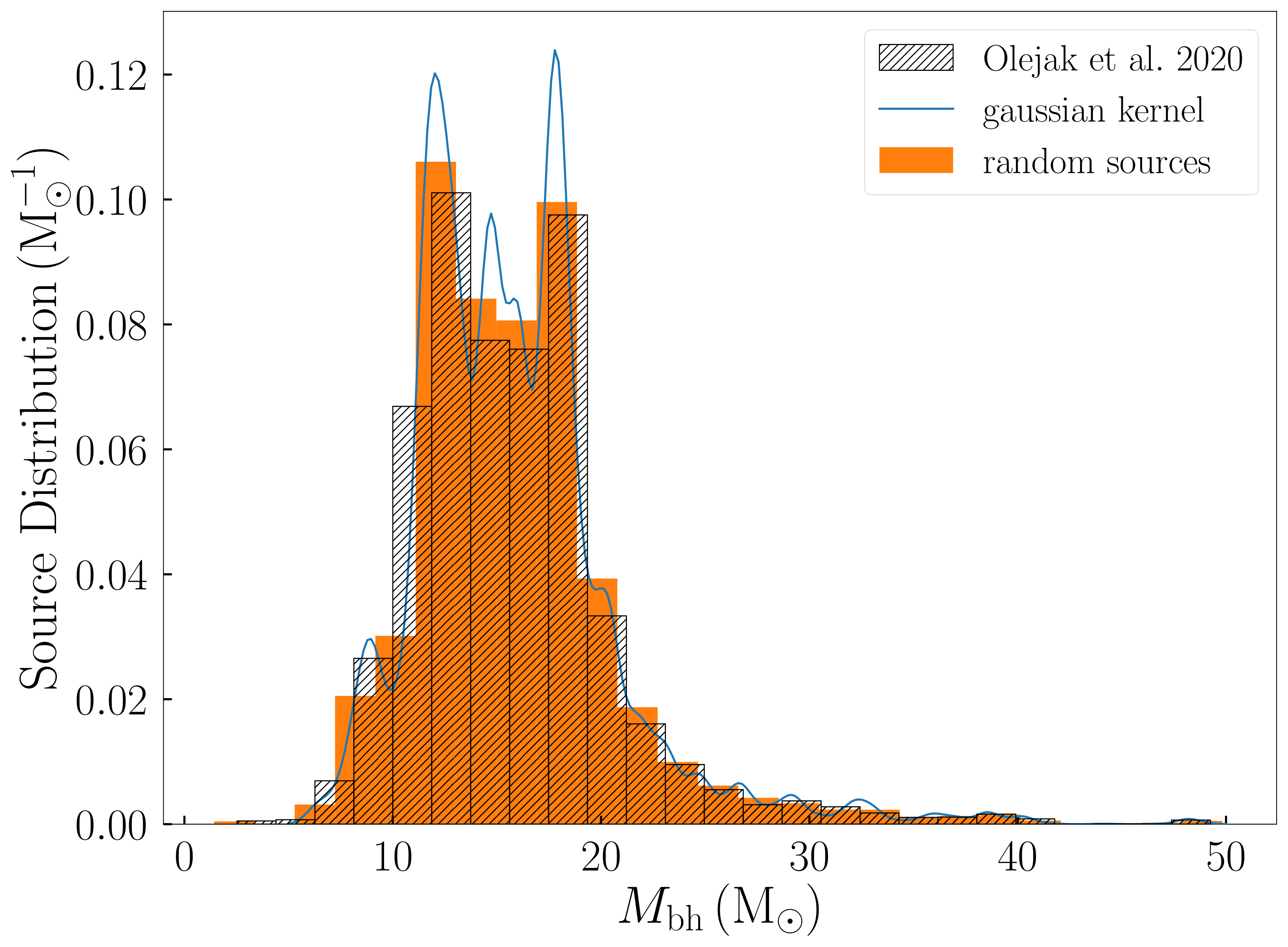}
    }
    \caption{Histogram of the masses of the BH of the q\bhs. The shaded grey region is adopted from \protect\cite{Olejak2019synthesis}, from which we derive the Gaussian kernel shown by the solid line. The shaded orange histogram corresponds to $10^4$ boxy bulge q\bhs (\textit{left panel}) and $1.2\times 10^5$ disc q\bhs (\textit{right panel}) used in this work. } 
    \label{fig: Mass distribution histogram}
\end{figure*}

The simultaneous radio to X-ray observations of binary systems have led to the identification of about 50 \bhs \citep{Corral_Santana_2016,Tetarenko2016}. The duty cycle of these sources is approximately 10~\%, that is, they spend most of their lifetime in quiescence. Following the work of \cite{Olejak2019synthesis}, we extracted information about a recent population synthesis analysis. More precisely, we separately studied three different regions of the Milky Way: a Lorimer-like population of q\bh in the Galactic disc \citep{Lorimer2006}, a further population in the boxy bulge, and a third distribution up to a few dozen parsec around the Galactic centre (GC). As the absolute size of each population, we adopted the following numbers: $10^3$ sources in the GC, $10^4$ in the boxy bulge, and an upper limit of $1.2\times 10^5$ sources in the disc (\citealt{Olejak2019synthesis} and see discussion below). We drew random values in a 3D grid to place the sources in the three different Galactic regions, namely, we adopted a Gaussian distribution for the radial distances of the GC sources with a mean value of 2\,pc and a standard deviation of 20\,pc \citep{Mori_2021}. For the boxy bulge, we followed the formula of \cite{cao2013new}, and for the disc sources, we adopted the Lorimer distribution of \cite{Lorimer2006}. 
We assumed that the Galactic disc is a 2D structure with a radius of 20\,kpc and a height of 2\,kpc above and below the Galactic plane, and the location of the Solar System at 8.3\,kpc from the GC. In Fig.~\ref{fig: distances and skymaps}, we show the sky map and overplot the distances of the sources of each individual population with the GC, the boxy bulge in the middle, and the Lorimer-like population in the lower panels. In the left panels, we show the sky maps in Galactic coordinates, and on the right, we plot the histograms of the distances of the produced q\bhs from the Sun.

In addition to the spatial positions, we also modelled the BH mass distribution. 
\cite{Olejak2019synthesis} presented the distribution of the mass of the BH in binaries for the Galactic disc and the Galactic halo.
In Fig.~\ref{fig: Mass distribution histogram}, we plot the histogram based on the supplementary material of \cite{Olejak2019synthesis}, for which we find a Gaussian kernel to use as probability function distribution (PDF). From this calculated PDF, we extracted $10^4$ and $1.2\times 10^5$ random values of the mass of the BH for the boxy bulge and the Lorimer-like distribution, respectively. 
We overplot the extracted random values as an orange-shaded histogram on the resulting distribution of \cite{Olejak2019synthesis}. 

A further important parameter of the q\bhs is the viewing angle, that is, the angle between the jet axis and the line of sight. We lack robust constraints and therefore used a uniform PDF between $1^{\circ}$ and $90^{\circ}$ to evaluate the viewing angle of each q\bh.

For each unique source, we assumed a set of parameters with the mass of the BH, the viewing angle, and the distance. Assuming that all q\bhs behave similarly to \src, we used these three quantities to rescale the emitted spectrum.

\begin{figure}
    \centering
        \includegraphics[width=1.\columnwidth]{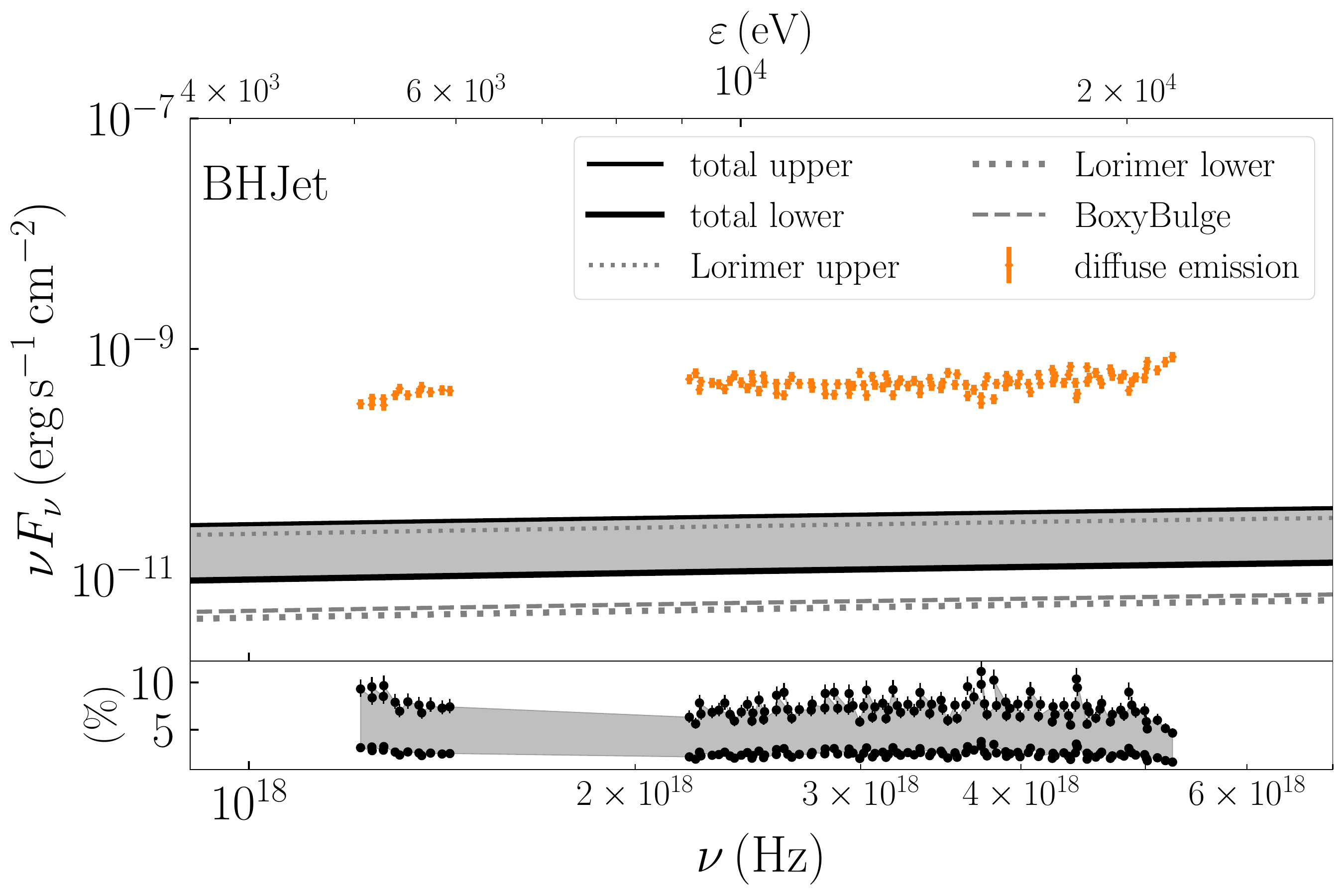}\\
        \includegraphics[width=1.\columnwidth]{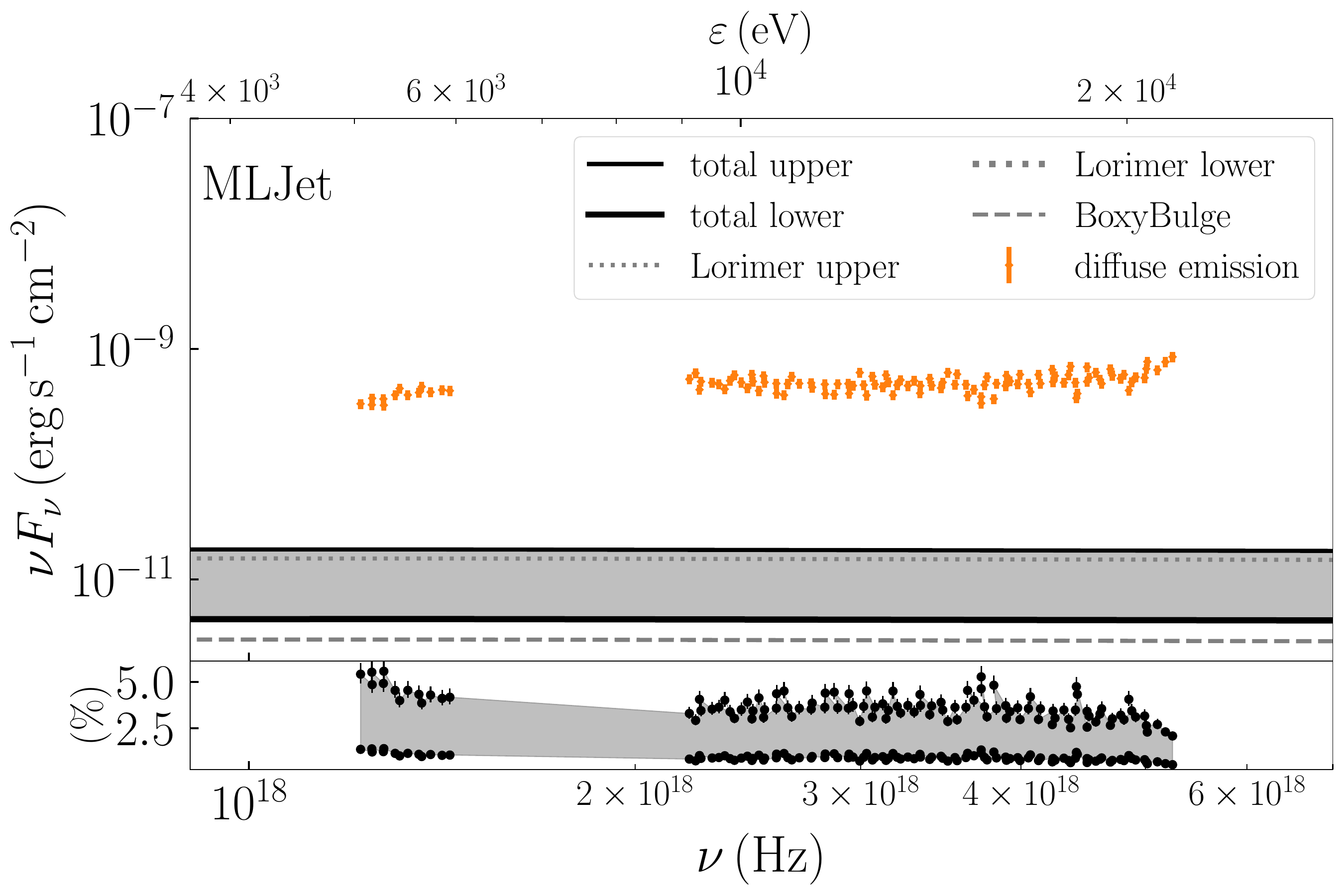}

    \caption{Contribution of the prompt emission of the Galactic q\bhs to the NuSTAR diffuse emission for $1^{\circ}\leq (l,b) \leq 3^{\circ}$ \protect\citep{perez2019galactic}. The upper panel shows the assumed model \blj, and the lower panel shows the \mlj. In both panels, we show the lower and upper limit of the Lorimer-like distribution ($1.2\times 10^4$ versus $1.2\times 10^5$ q\bhs) and the $10^3$ boxy bulge-like sources based on \protect\citep{Olejak2019synthesis}. The sum of the boxy bulge and the Lorimer-like distribution leads to the total contribution, which we plot with the thin solid line for the upper limit and the thick solid line for the lower limit. The shaded grey region corresponds to the predicted contribution of these q\bhs to the observed diffuse emission. The insets in both panels show the percentage contribution per energy bin.
    }
    \label{fig: spectral contribution of qbhxrbs Nustar}
\end{figure}

\begin{figure}
    \centering
        \includegraphics[width=1.\columnwidth]{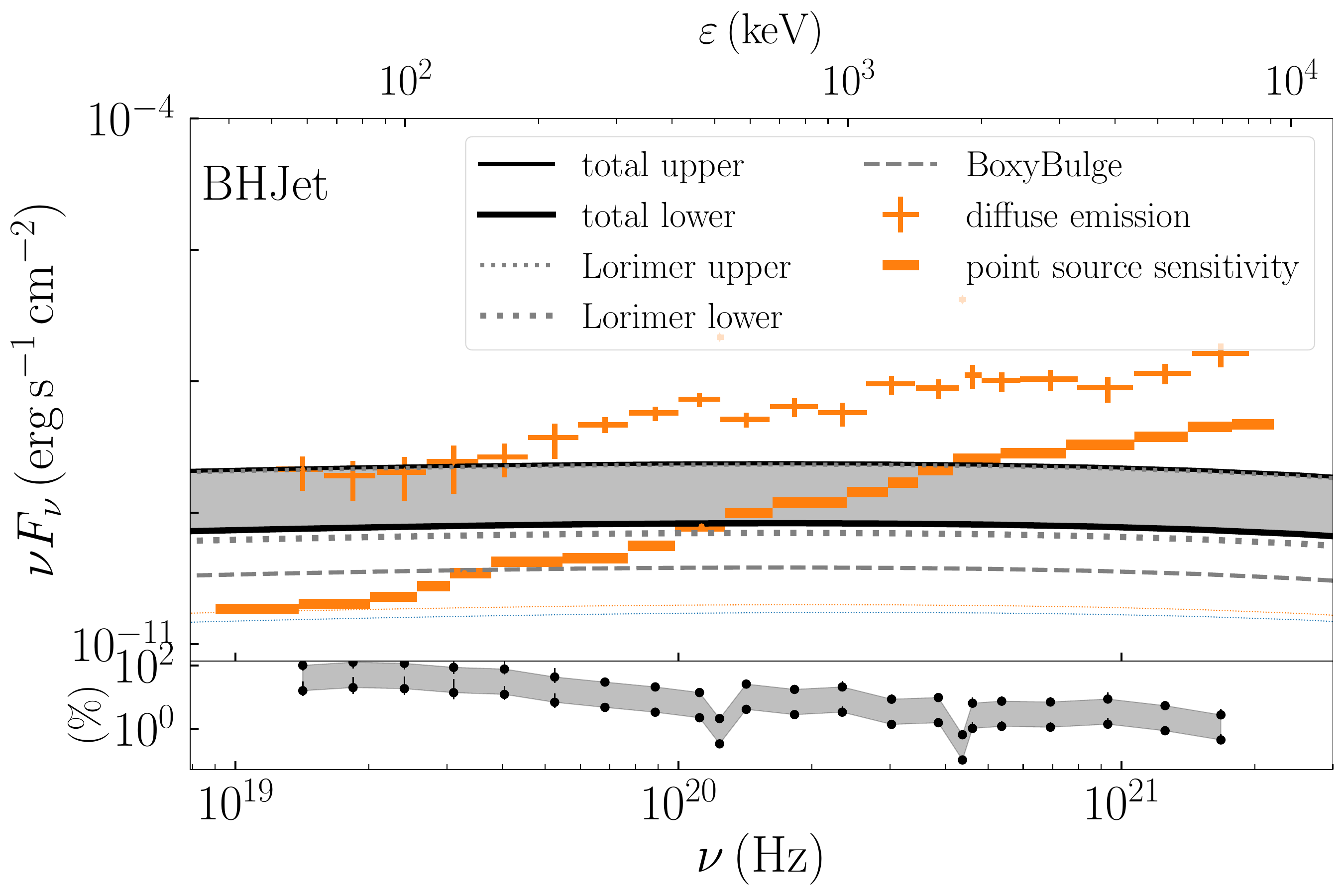}\\
        \includegraphics[width=1.\columnwidth]{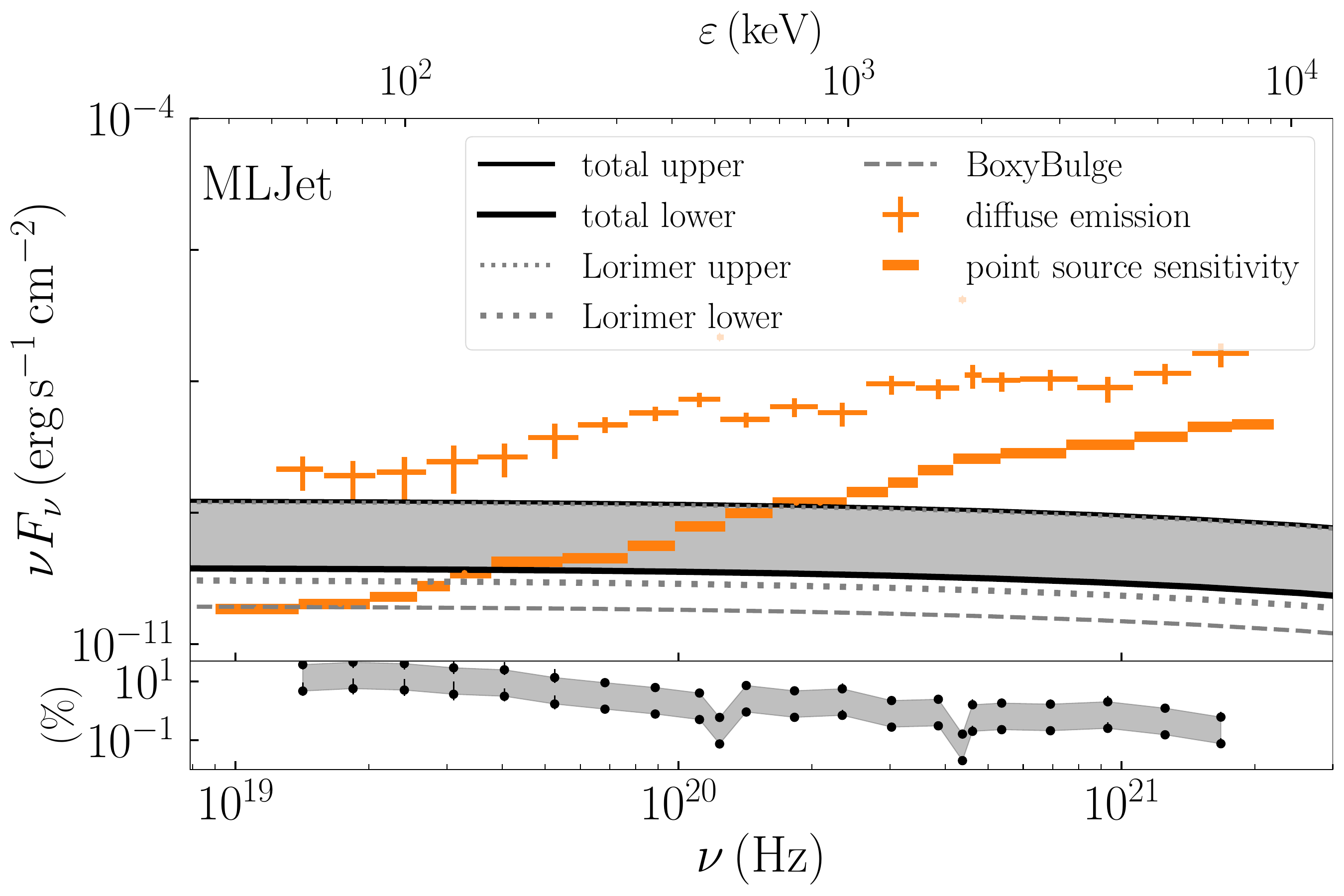}
        
    \caption{ Similar to Fig.~\ref{fig: spectral contribution of qbhxrbs Nustar}, but for the energy band covered by \integral for $(|l|, |b|) \leq (47.5^{\circ}, 47.5^\circ)$ \citep{berteaud2022strong}. We overplot the point source sensitivity of \spi on the diffuse emission, as indicated in the legend \protect\citep{Roques2003spi}, and compare it to individual sources that contribute to the total emission (solid coloured lines for individual sources). We show only sources that lie above the instrument sensitivity. No sources lie above the threshold for \mlj.
    }
    \label{fig: spectral contribution of qbhxrbs Integral}
\end{figure}

\begin{figure}
    \centering
        \includegraphics[width=1.\columnwidth]{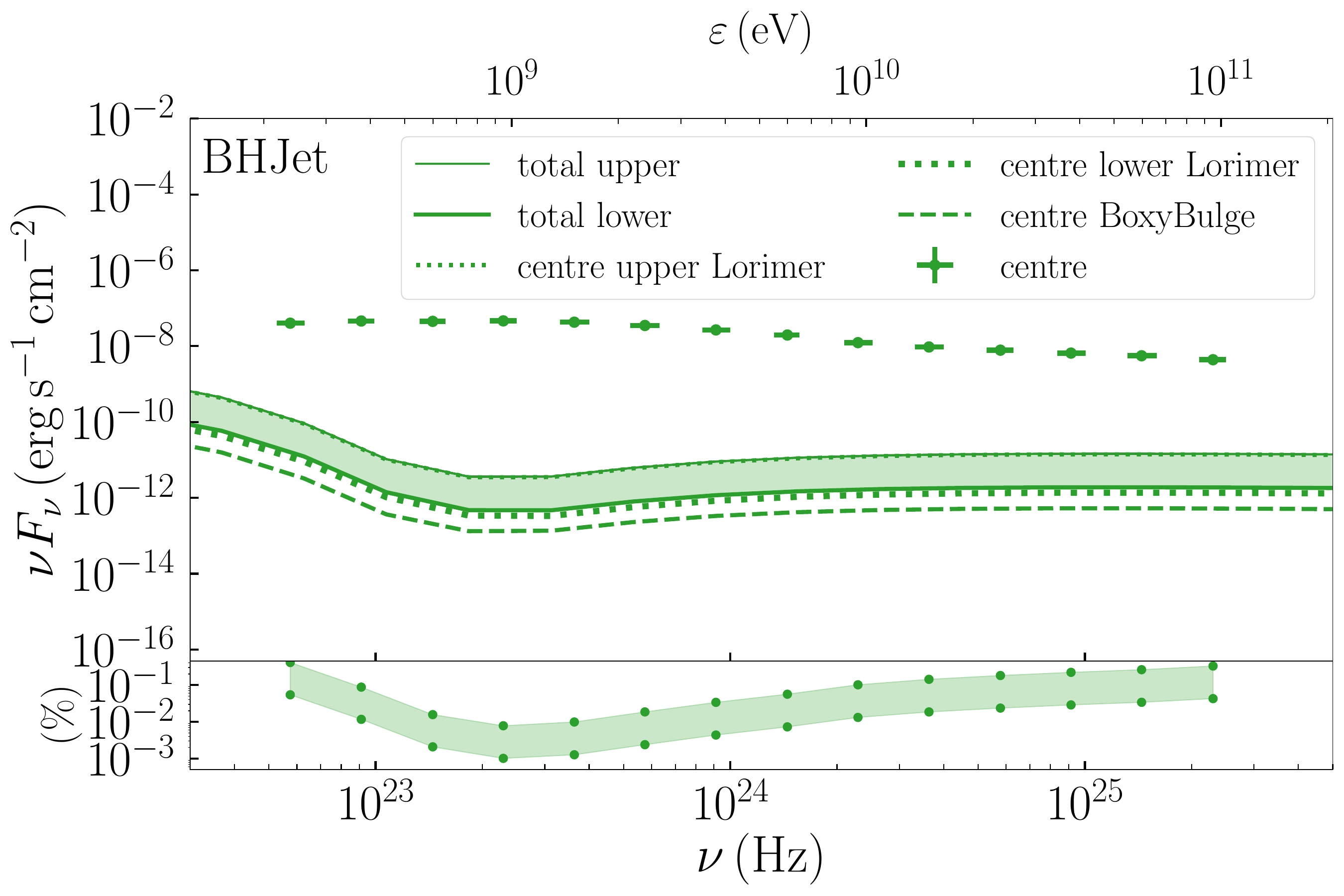}
        \includegraphics[width=1.\columnwidth]{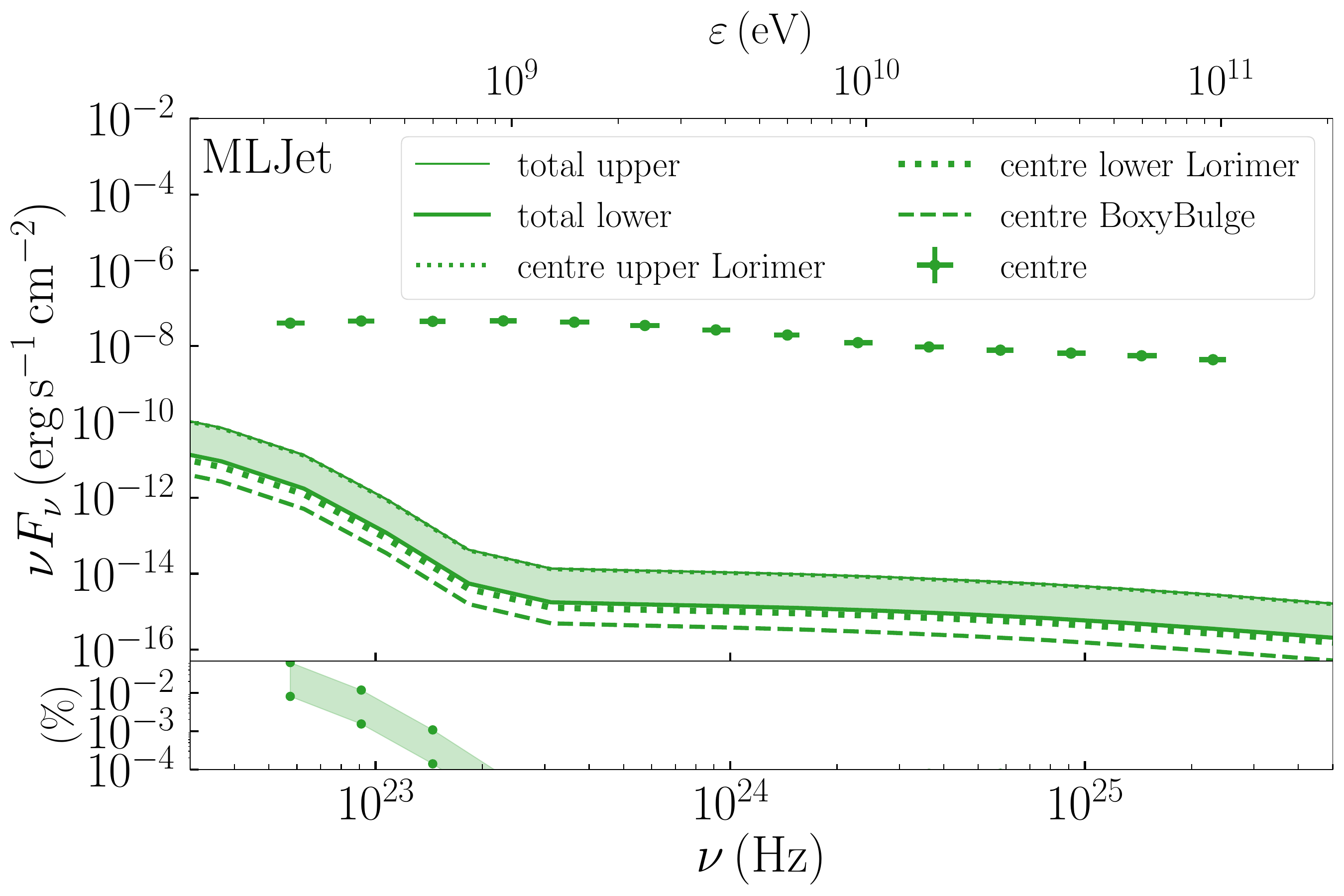}

    \caption{Similar to Fig.~\ref{fig: spectral contribution of qbhxrbs Nustar}, but for the energy band covered by \fermi, and only for the region around the GC with $|l| \leq 80^{\circ}$ and $|b| \leq 8^{\circ}$.  }
    \label{fig: spectral contribution of qbhxrbs Fermi center}
\end{figure}

\begin{figure*}
    \centering
	\subfigure{
        \includegraphics[width=0.975\columnwidth]{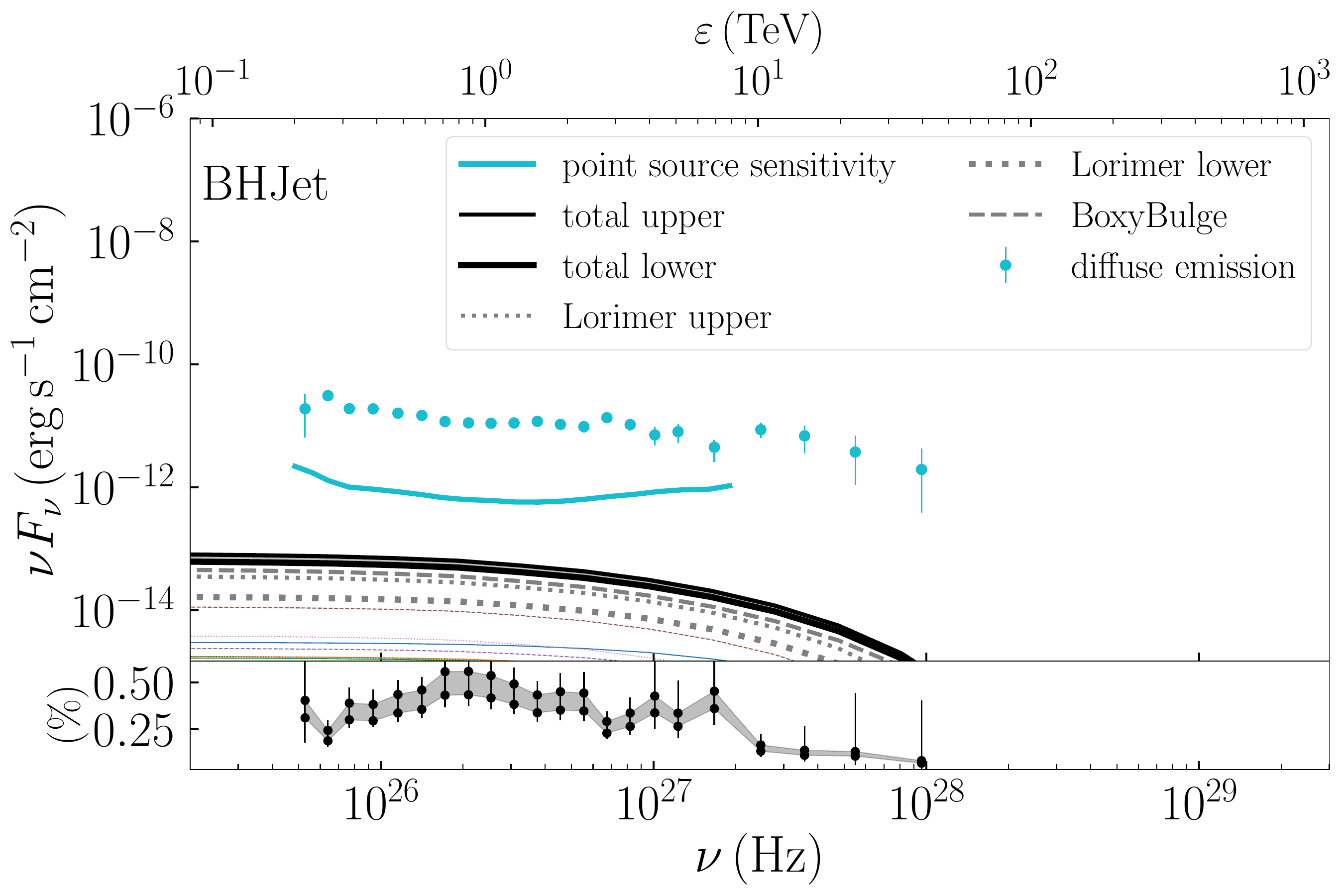}
    }
    \subfigure{
        \includegraphics[width=0.975\columnwidth]{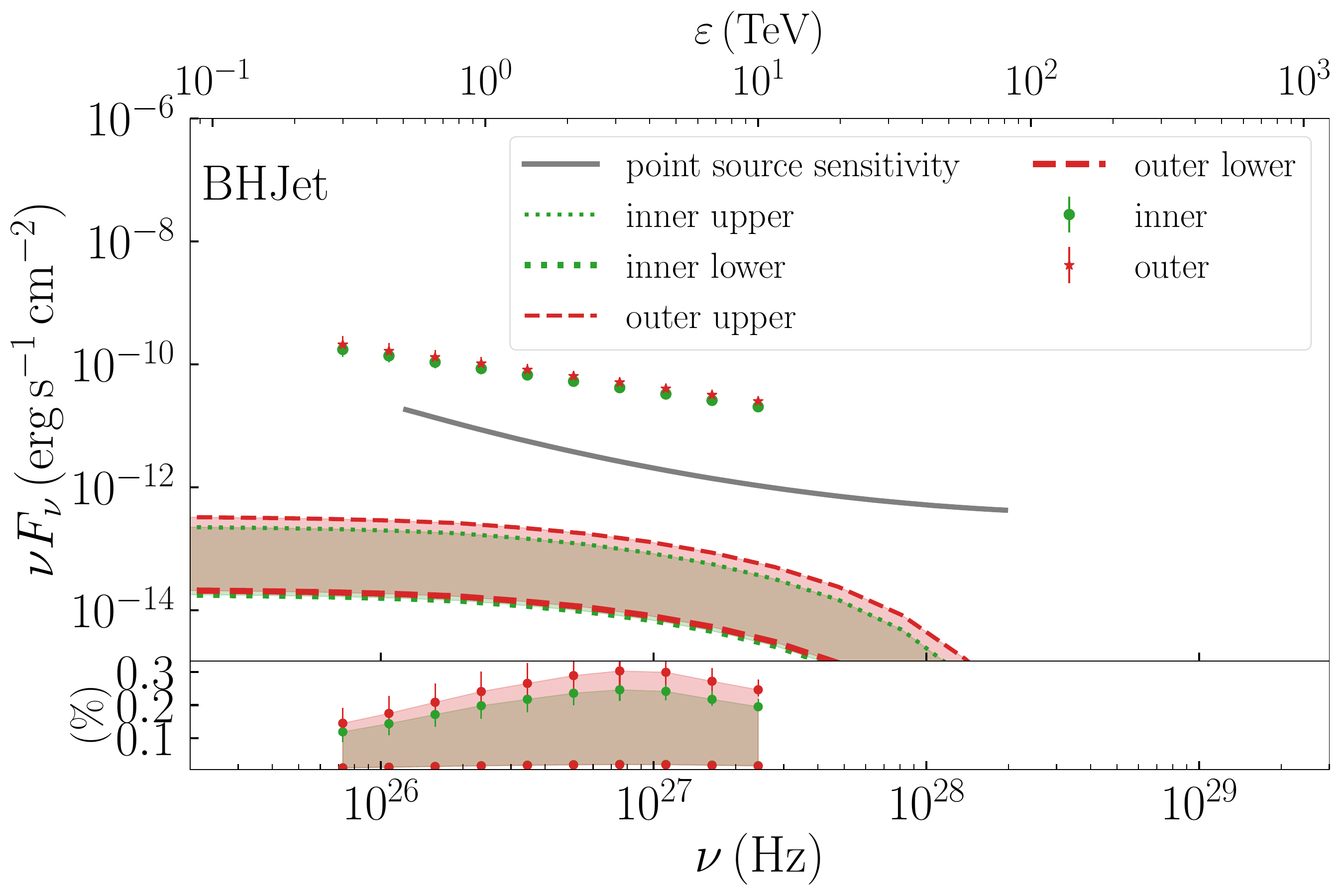}
    } 
	\subfigure{
        \includegraphics[width=0.975\columnwidth]{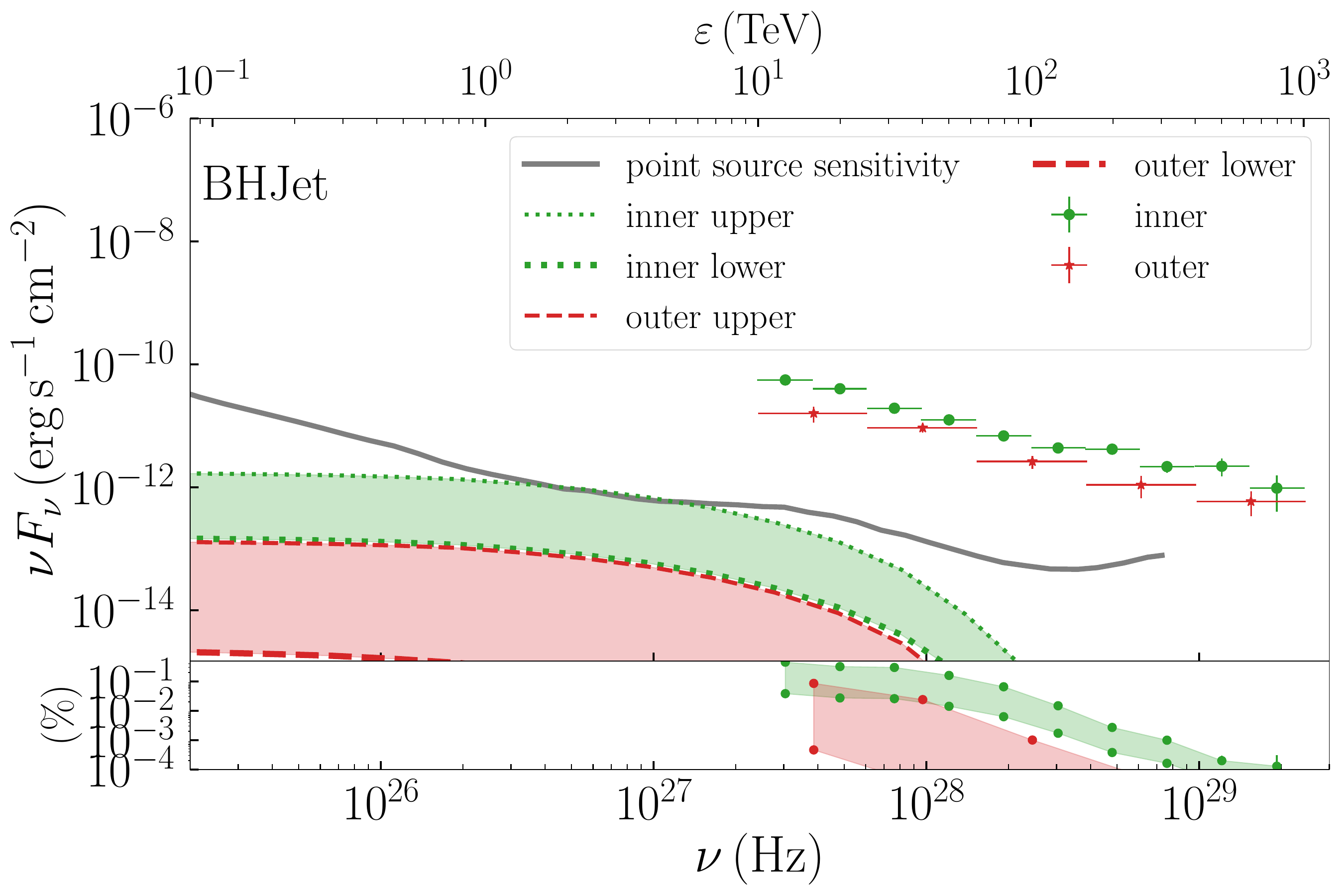}
    }
	\subfigure{
        \includegraphics[width=0.975\columnwidth]{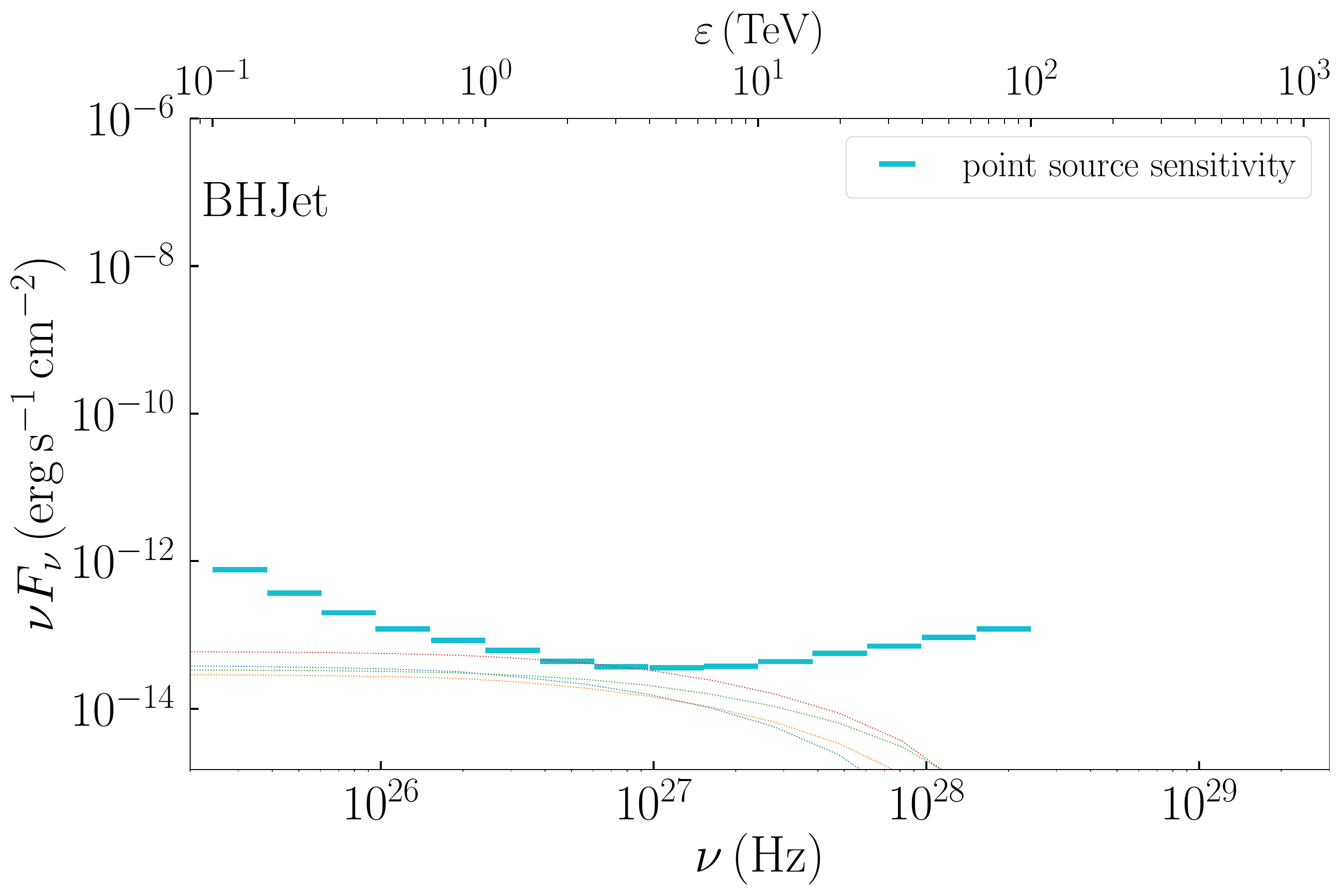}
    }
    \caption{Similar to Fig.~\ref{fig: spectral contribution of qbhxrbs Nustar}, but for the highest-energy regime of TeV compared to the \hess (\textit{top left}), HAWC (\textit{top right}), LHAASO (\textit{bottom left}), and CTAO (\textit{bottom rigth}) facilities. For \hess, HAWC, and LHAASO, we show with solid coloured lines the individual sources that exceed $10^{-3}$ of the point source sensitivity, and for CTAO, we show the sources that exceed one-third of the predicted point source sensitivity. For all panels, we only show the case of \blj because the contribution of sources that follow \mlj is even smaller than pictured here. The inner (outer) region for the case of HAWC is $43^\circ \leq l \leq 73^\circ$ and $|b|\leq 2^\circ$ ($|b|\leq 4^\circ $), and the inner (outer) region for the case of LHAASO is $15^\circ \leq l \leq 125 ^\circ $ ($125^\circ \leq l \leq 235 ^\circ$) and $|b|\leq 5^\circ$. For CTAO, the sources lie within $|l| \leq 60^\circ$ and $|b| \leq 3^\circ$ \citep{eckner2023detecting}. All the panels have the same axes for comparison. } 
    \label{fig: spectral contribution of qbhxrbs TeV gamma rays  }
\end{figure*}

\subsection{Multi-wavelength prompt emission}
In Figs.~\ref{fig: spectral contribution of qbhxrbs Nustar}--\ref{fig: spectral contribution of qbhxrbs TeV gamma rays  }, we plot the cumulative emission of all these q\bhs in different energy bands from keV X~rays to TeV \grs. More precisely, we used the detected diffuse emission from the following instruments to investigate the contribution of q\bhs: NuSTAR in the keV band, \integral in the $\sim$MeV X~rays, \fermi in the GeV \grs, and \hess, the High-Altitude Water Cherenkov Observatory (HAWC), LHAASO, and CTAO for TeV \grs. Moreover, we compared individual sources to the instrument sensitivity to discuss their potential detection.

We used the NuSTAR Galactic diffuse emission from \cite{perez2019galactic} for a region of $3^{\circ}$ around the GC. For the case of \integral, we adopted the analysis of \cite{berteaud2022strong}  for $\lvert \rm {l}\lvert \leq 47.5^\circ$ 
and $\lvert \rm {b}\lvert \leq 47.5^\circ$ for the diffuse emission and the sensitivity from \cite{Roques2003spi}. The \fermi data for different regions in the sky map were taken from \cite{Ackermann_2012} for four different regions: the Galactic disc with $\lvert \rm {l}\lvert \leq 80^\circ$ 
and $\lvert \rm {b}\lvert \leq 8^\circ$, the intermediate disc for $\lvert \rm {l}\lvert \leq 180^\circ$ 
and $ 10^{\circ} \leq \lvert \rm {b}\lvert \leq 20^\circ$, the ridge with $80^{\circ} \leq \lvert \rm {l}\lvert \leq 180^\circ$ and $\lvert \rm {b}\lvert \leq 8^\circ$, and the off-plane region with $\lvert \rm {l}\lvert \leq 180^\circ$ 
and $ 8^{\circ} \leq \lvert \rm {b}\lvert \leq 90^\circ$. 
The \hess diffuse emission was taken from \cite{HESS2018GalacticCenter} for a region of $\lvert \rm {l}\lvert \leq 1^\circ$ and $\lvert \rm {b}\lvert \leq 5^\circ$. 
The best fit of the HAWC Galactic diffuse emission for $43^{\circ} \leq \rm {l} \leq  73^{\circ} $ and two regions on the sky map with $\lvert \rm{b} \lvert \leq 2^\circ $ for the innermost one, and $\lvert \rm{b} \lvert  \leq 4 ^{\circ} $ for the outermost one,
are $8.89\pm 0.37 ^{-0.70}_{+0.48} \times \left( E/7.0\,\rm{TeV} \right)^{-2.612\pm 0.030 ^{-0.036}_{+0.015}}$ and $5.45\pm 0.25 ^{-0.44}_{+0.38}  \times \left( E/7.0\,\rm{TeV} \right)^{-2.604\pm 0.034 ^{-0.037}_{+0.012}}$ in units of $10^{-12}\, \rm{TeV^{-1}\,s^{-1}\,cm^{-2}\,sr^{-1}}$, respectively \citep{nayerhoda2021HAWCgalactic}. We fixed a probable typo in the provided units. 
The HAWC sensitivity after 507~days was derived from \cite{abeysekara2017HAWCobservation}. 
For the $\sim$PeV regime and the LHAASO collaboration, 
there are two sky map regions with $\lvert \rm{b} \lvert \leq 5^{\circ}$, the innermost map with $15^{\circ}\leq \rm{l} \leq 125 ^\circ $ and the outermost map with  $125^{\circ}\le \rm{l} \leq 235 ^\circ $ \citep{LHAASO2023Galactic}. Finally, we adopted the CTAO simulated sensitivity for $\lvert \rm {l}\lvert \leq 60^\circ$ 
and $\lvert \rm {b}\lvert \leq 3^\circ$ from \cite{eckner2023detecting}. For each different energy regime, we only used sources that lay within the aforementioned regions to calculate the cumulative emission.

In Fig.~\ref{fig: spectral contribution of qbhxrbs Nustar}, we show the predicted cumulative emission of the Lorimer distribution and the boxy bulge in the keV spectrum. For simplicity, we neglected the GC because it almost does not contribute at all. In the upper panel, we show the cumulative emission assuming that all the q\bhs obey the \blj model, and in the lower panel, we show the \mlj. 
As described above, the $1.2 \times 10^5$ sources of \cite{Olejak2019synthesis} host a BH with a non-degenerate companion, but not necessarily all of these BHs accrete matter from the companion at the same time. We therefore used this value as an upper limit (the densely dotted line labelled `Lorimer upper'), and we combined it with a lower limit assuming that 10\% of these sources accrete at a time (see the more detailed discussion below). Summing the boxy bulge with either the upper or lower limit of the Lorimer yields the total emission in the keV X-ray band between some upper and some lower limits (total upper and total lower, respectively). The shaded grey region therefore shows the total predicted emission from the Galactic q\bhs between the upper limit of the Lorimer and some lower case of 10\%. In the insets of the upper and lower panels, we show the percentage of the potential contribution of the q\bhs to the NuSTAR diffuse emission. For both jet models and depending on the number of disc sources, the contribution might be of about a few to 20\%. 

In Fig.~\ref{fig: spectral contribution of qbhxrbs Integral}, we plot the contribution of the q\bhs to the \integral diffuse emission (orange crosses) following the above pattern. The contribution of q\bhs to this energy band is larger; it reaches values of almost 100\% in the $\sim 100$\,keV to a few percent in the regions of dozens of MeV. In the same plots, we include the sensitivity of \spi for point sources from \cite{Roques2003spi} to compare to individual sources. 
For the case of \blj, we find that a few sources might exceed the sensitivity in the first several energy bins in the $\sim$40-100\,keV. 

In Fig.~\ref{fig: spectral contribution of qbhxrbs Fermi center}, we plot the predicted contribution of q\bhs to the GeV \grs compared to the \fermi detection of the extended region around the GC. We again used the same pattern as above to derive the upper limit of the total emission assuming an upper limit for the Lorimer-like disc distribution, and a smaller fraction of this at 10\%. The overall contribution drops to less than 0.1\% in these energy bins, and for the case of the \blj, the contribution drops below $10^{-4}$\% at energies beyond GeV. For \mlj on the other hand, the contribution can reach about 0.1\% not only in the GeV energies, but also in the 100\,GeV. In Appendix~\ref{app: fermi diffuse emission for intermediate, ridge and off plane}, we show the predicted contribution of q\bhs at intermediate Galactic latitudes, the ridge and the off plane for \blj and \mlj. We see no more than 0.1\% in the 1\,GeV and 100\,GeV energy bins, similar to the GC.

Finally, in Fig.~\ref{fig: spectral contribution of qbhxrbs TeV gamma rays  } we combine the three operating TeV facilities of \hess, HAWC, and LHAASO with the under-construction next-generation facility of CTAO. q\bhs can contribute up to 0.5\% of the 1-10\,TeV diffuse emission detected by \hess, but no individual sources are expected to be detected by \hess (no individual sources exceed the plotted sensitivity). The q\bhs of the ridge, that is, only sources from the disc, can contribute up to 0.3\% in the TeV diffuse emission detected by HAWC, and up to 0.1\% in the 10\,TeV emission detected by LHAASO. For LHAASO, and the outer region, in particular, where $125^{\circ} \leq l \leq 235^{\circ} $, the contribution drops as a function of energy and falls below $10^{-4}$\%. Finally, in Fig.~\ref{fig: spectral contribution of qbhxrbs TeV gamma rays  }, we compare some individual sources that we find that may produce significant TeV emission close to $10^{-13}\,\rm erg\, s^{-1}\, cm^{-2}$ with the predicted CTAO sensitivity for point source detection in the Galactic plane. In Fig.~\ref{fig: spectral contribution of qbhxrbs TeV gamma rays  }, we only include the predicted emission from q\bhs using the \blj jet model because the case of \mlj yields some TeV emission that is significantly lower, and the overall contribution therefore drops significantly to even lower than $10^{-4}$\%. We adopted this threshold here.

\subsection{Cosmic-ray fluxes and multi-wavelength secondary emission}\label{Sec: DRAGON results for propagating CRs}

We have only accounted for the intrinsic emission from the jets of q\bhs for either \blj or \mlj above. This intrinsic radiation is the product of the non-thermal CRs that are accelerated within the jets. When these CRs, or more precisely, a fraction of these CRs, escape from the acceleration sites, they propagate in the Milky Way. Using numerical means to follow this propagation, we estimated the CR flux detected here on Earth, and we compared it to the detected spectrum. In particular, we used \dragon, a publicly available simulator of the Galactic propagation. We assumed that the CR sources are the same as above, namely a Gaussian population around the GC, a boxy bulge, and a Lorimer-like disc distribution. The normalisation of each distribution is thoroughly explained in Appendix~\ref{app: the setup for dragon and the distributiions}. Each q\bh ejects some CRs in the Milky Way that follow a power law in energies similar to the intrinsic power law, and they carry some total power that we show in Table~\ref{tab: particle energetics}. Moreover, we calculated the maximum of the CR energy to be about 20\,TeV (see Table~\ref{tab: particle energetics}). As we show in Appendix~\ref{app: CR proton and electron spectra}, the contribution of the propagated protons is smaller than $10^{-10}$, and it is about $10^{-12}$ for electrons. 

\begin{table}
	\centering
        \setlength{\tabcolsep}{8pt} 
	\renewcommand{\arraystretch}{1.5} 
	\caption{Calculated power carried by the non-thermal protons (p) and electrons (e) and their maximum energy for the two jet models.}
	\label{tab: particle energetics}
	\begin{tabular}{l|cc} 
		\hline
		  & \blj  &  \mlj \\
		\hline
              $P_p\, (10^{33}\,\rm erg\,s^{-1})$& $310$ & $4$   \\
              $P_e\, (10^{33}\,\rm erg\,s^{-1})$& $17$ & $11$ \\
              $E_p$ (TeV)& $17$& $30  $ \\
              $E_e$ (TeV)& $20$& $13 $ \\	    
		\hline
	\end{tabular}
\end{table}


\section{Discussion}\label{Sec: discussion}

\subsection{\blj vs \mlj for \src}

To finally determine the exact number of q\bhs in the Milky Way and their role in the multi-wavelength emission, we need to investigate the physics of individual sources better, such as \src. With the \blj and \mlj jet models, we captured the broadband electromagnetic picture from radio to X-rays. More precisely, the thermal synchrotron emission from the jet base can explain the IR emission along with the stellar emission of the companion, in agreement with \cite{dinccer2017multiwavength}. The synchrotron optically thick emission connects the radio to the X-ray emission, similar to various \bhs in outburst \citep{Merloni2003fundamental,Falcke2004FP,Plotkin2014Constraints}. In both cases, however, the IC contributes equally, and it is therefore impossible to distinguish the two components with a spectral fitting alone. Further means are needed, such as a timing analysis. Nonetheless, we were able to constrain the jet dynamics to predict the hard X-ray and \gr emission. 
In both scenarios and assuming an efficient particle acceleration, the optically thick synchrotron emission reaches the MeV bands that emerge from a cooled population of non-thermal electrons. 
The distribution of these electrons is the convolution of a thermal Maxwell-J\"uttner and a non-thermal tail that expands to energies of about 10~TeV (see Table~\ref{tab: particle energetics}) for a power-law index close to 2 (see Table~\ref{table: free parameters}). This distribution is similar to the one plotted in Fig.~6 of \kgx, but for a softer index. The radiation footprint of this leptonic population to dozens of TeV energies is the flat spectrum in the keV-MeV regime in a $\nu F_{\nu} (\nu)$ plot (see the Syn, $z>z_{\rm diss}$ component of Fig.~\ref{fig: hadronic MW SED synchrotron dominated X rays}). This sub-MeV emission is hard to detect with either \integral or \fermi under these jet conditions. 

The main differences between the two jet models are in the GeV and beyond spectrum and in the jet composition. Starting from the latter, for \blj we assumed an equal number density of electrons and protons all along the jet. For \mlj, on the other hand, as we describe in Sec.~\ref{Sec: Jet model}, the jets are launched with $\eta$ pairs of electrons and positrons with respect to pairs of electrons and protons. For \src, we used this as a free parameter, and we found that $\sim$200 pairs per proton are needed to explain the radio to X-ray emission. This difference in the jet composition leads to different spectral components in the TeV band. In particular, the neutral pion decay from pp interactions is reduced in the case of \mlj because fewer targets reside in the jets that allow for an inelastic collision, but more target electrons exist, which enable a stronger IC component (stronger than \blj, but still not strong enough to increase the GeV flux significantly). This more physically motivated proton power allows for better constraints in terms of jet energetics, but unfortunately, it does not lead to stronger non-thermal \gr emission.

\subsection{Population of \bhs in the disc}

According to \cite{Olejak2019synthesis}, $1.2\times 10^5$ q\bhs may reside in the Galactic disc, but this number is very uncertain. 
However, only a fraction of these sources are currently expected to accrete and launch jets simultaneously. Based on the radio detection of another q\bh, namely VLA~J2130+12, 
\cite{tetarenko2016first} estimated that  $2.6\times 10^4-1.7\times 10^8$ objects may exist in the Milky Way at the moment. The lower limit is of the same order of magnitude as the value we used as a lower limit of the current q\bhs (for simplicity, we used $1.2\times 10^4$ sources, that is, 10\% of the total population). This number of q\bhs might agree with the more recent observations of \textit{Gaia}, which  detected approximately $6\times 10^3$ ellipsoidal systems, a few percent of which harbour a compact object \citep{Gomel2023gaiaEllipsoidal}. \textit{Gaia} detected only nearby binaries, however, and only those with an orbital period shorter than a few days can be claimed to be binaries. 

The estimated lower limit of \cite{tetarenko2016first} is consistent with the upper limit of theoretical predictions for the number of \bhs in the Milky Way based on population synthesis (see e.g. \citealt{Romani1992,Portegies_Zwart1997,Kalogera_1998,Pfahl_2003,yungelson2006origin,Kiel2006}). 
Nonetheless, in the population synthesis studies, the fraction of accreting \bhs is typically $\sim 0.01$~\% within a Hubble time \citep{yungelson2006origin}. The more recent analysis by \cite{Wiktorowicz2019population} showed that $\sim 10^4$ \bhs are expected to fill their Roche lobe. This value is only 1\% of the total population of sources. 

The estimated upper limit of \cite{tetarenko2016first} is approximately two orders of magnitude above the upper limit we used. Figs.~\ref{fig: spectral contribution of qbhxrbs Nustar} and \ref{fig: spectral contribution of qbhxrbs Integral} showed that the X-ray diffuse emission does not allow for more than $10^5$ jetted q\bhs at the same time. A further comparison, however, might not lead to concrete conclusions because our main assumption that all these $1.2\times 10^4-1.2\times 10^5$ qBHXBs launch jets of approximately the same power cannot rule out the existence of even more qBHXBs that lack any jet emission. Consequently, it is worthwhile to revisit the contradiction between the theoretical prediction of $\leq 10^4$ and the observational evidence for $ \geq 10^4$ so that we can better constrain the contribution of (q)\bhs to the observed diffuse Galactic radio to \gr emission detected, and vice versa, the further observational investigation of (q)\bhs can help us to better constrain the contribution of these sources to the Galactic diffuse emission.

\subsection{X-ray to \gr contribution to the diffuse emission}

The diffuse X-ray emission detected by NuSTAR is mainly up to 90\% due to cataclysmic variables \citep{perez2019galactic}. Some 10\% of this emission, however, is due to unresolved Galactic sources. We found that q\bhs that launch jets can contribute up to 10\% depending on the jet model. The contribution can be reduced to a few percent because the number of accreting q\bhs drops to 10\%, that is, about $1.2\times 10^4$ sources in the Galactic disc. 

In the more energetic X-ray regime, and in particular, in the energy window observed by \integral, we find that the q\bhs can fully explain the $\sim 100$\,keV band and up to a few percent to the regime of some dozen keV. It is worth mentioning that in the X-ray analysis of the \integral data, the contribution of the unresolved sources is commonly assumed to follow a power law in energy with an exponential cutoff \citep[see, e.g.][]{berteaud2022strong}, but based on this analysis, the contribution of q\bhs is energy dependent and decreases with energy. This energy dependence of the contribution is due to radiatively cooled electrons that are accelerated in the q\bhs jets. 
For a less efficient particle acceleration in which the particles escape from the acceleration sites long before they reach their theoretical maximum energy, the X-ray spectrum would show a cutoff at much lower energies than some dozen MeV, but still in the X-ray band. The contribution of q\bhs to the high-energy tail may hence drop, unlike in the lower regime at 100\, keV.

When we compare the \integral sensitivity for individual sources to the predicted emission of some individual q\bhs that contribute the most to the diffuse emission in the softer regime of the MeV X-ray spectrum, we expect one source at most to be detected in the 30-100\,keV regime for the case of \blj . For a more accurate prediction of this value, we performed 250 realisations of our simulations to find othat $0.4\pm 0.5$ sources are expected to be detected by \integral on average. 
This value indeed agrees well with the number of \bhs detected by \integral so far, but no q\bh has been detected to date.

In the higher-energy regime of \grs, the contribution of q\bhs is smaller than 1\% in the entire spectrum from GeV to TeV energies. The CTAO, however, is expected to be ten times more sensitive than current facilities. To predict numbers more reliably for this energy regime, we realised 250 simulations, according to which, $1\pm 0.5$ sources are expected to be detected by CTAO. It is worth remarking that these sources are merely in the quiescence and not during outburst, when stronger TeV emission is expected (see e.g. \citealt{Kantzas2023MassLoading}). 
For these sources that exceed the CTAO sensitivity in each realisation of the simulations, the CTAO will be able to detect sources that are located closer than $\sim 2\,$kpc (with an average distance of $0.9\pm 0.4\,$kpc) and have a viewing angle smaller than $17^{\circ}$ (with an average viewing angle of $11\pm 2 ^{\circ}$). The majority of the \bhs known so far are beyond 1\,kpc, and the nearest system lies at 0.5\,kpc \citep{El-Badry2023Sunlike,Chakrabarti_2023Noninteracting}. Only the viewing angle of MAXI~J1836--194 is smaller than 15$^\circ$ and located farther away than 4\,kpc \citep{Russell2014J1836}. We obtained these results under the assumption that all q\bhs follow the spectral behaviour of \src because the number of q\bhs with good-quality multi-wavelength data is still small. If CTAO can indeed detect TeV emission from individual q\bhs, this will allow us to study a new part of the parameter space.

\subsection{Radio counterparts and predictions for future facilities}
\begin{figure}
    \centering
        \includegraphics[width=1.\columnwidth]{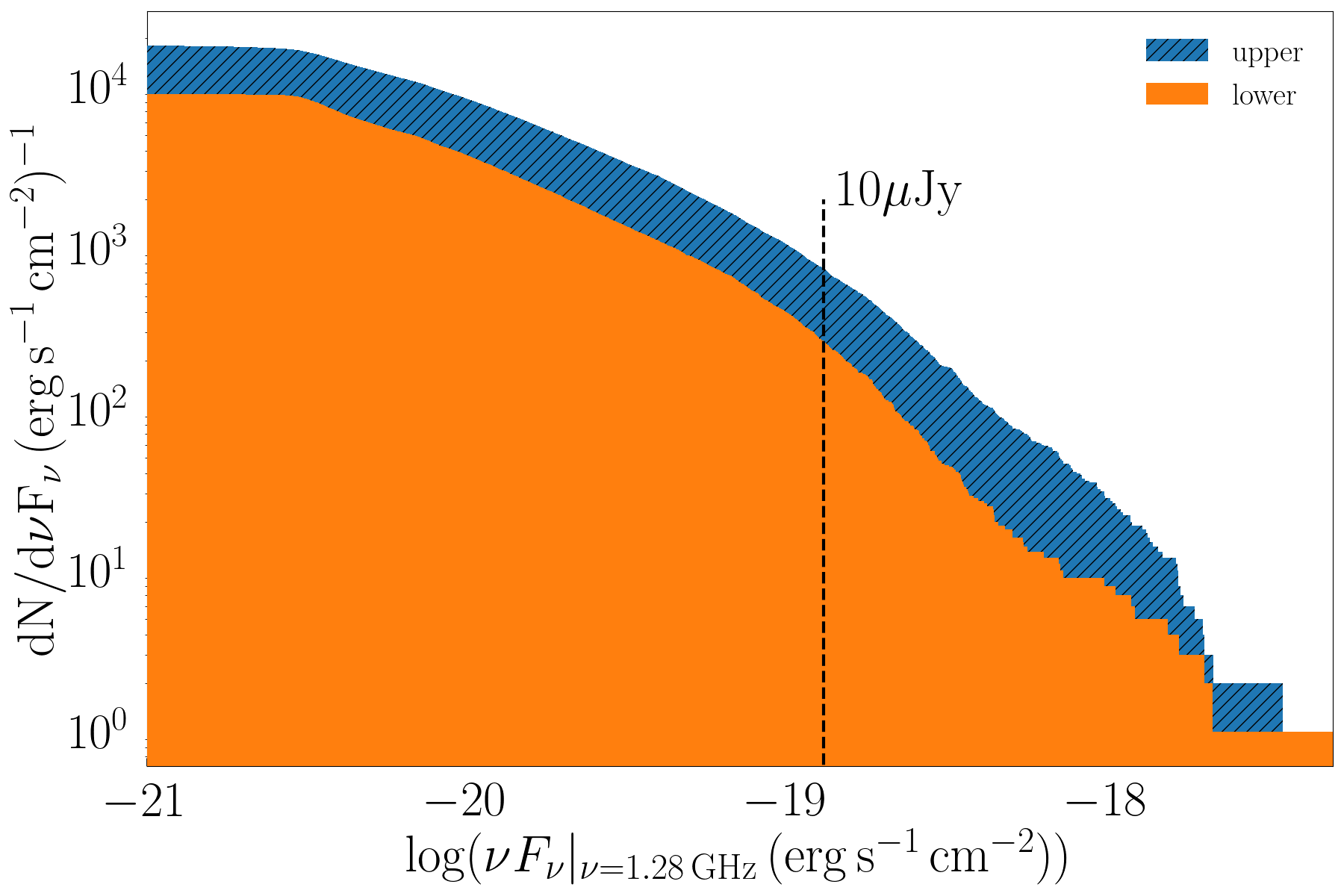}
    \caption{Cumulative distribution of sources per energy flux detected at a characteristic radio frequency of 1.28\,GHz. The two histograms correspond to the two assumptions for an upper or lower limit of q\bhs populating a Lorimer-like distribution in the disc. We show for comparison the detection threshold of radio sources with current facilities, such as MeerKAT and VLA ($\sim 10\mu$Jy).}
    \label{fig: radio counterpart luminosity flux}
\end{figure}

The radio counterpart of q\bhs, such as the one detected by the VLA from \src \citep{dinccer2017multiwavength}, is a key element to prove the jet emission, particularly in the case of a flat radio spectrum \citep{blandford1979relativistic}. Current facilities such as the VLA and MeerKAT can detect sources with a radio intensity lower than 100$\mu$Jy and can sometimes reach values close to 10$\mu$Jy \citep{Driessen2019firstMeerKAT,Goedhart2024SARAO}. In the future, with ngVLA \citep{butler2019ngvla} and the full array of the SKA \citep{braun2017anticipated}, the threshold for a detection may drop to $\sim \mu$Jy. 
In Fig.~\ref{fig: radio counterpart luminosity flux}, we plot the cumulative distribution of sources per energy flux at $1.28\,GHz$. The two different histograms correspond to the assumption of the upper/lower Lorimer distribution, similar to what we described above. For the current threshold of $\sim 10\mu$Jy, we may be able to detect up to a few hundred sources, depending on the number of jetted q\bhs (lower and upper limits, respectively). Simultaneous multi-wavelength observations in the optical and/or the X-rays, as well as a timing analysis that indicates periodicity, are necessary to identify these radio sources as (q)\bhs \citep[see e.g.][]{Driessen2019firstMeerKAT}.

\subsection{Neutrino counterpart}

\begin{figure}
    \centering
        \includegraphics[width=1.\columnwidth]{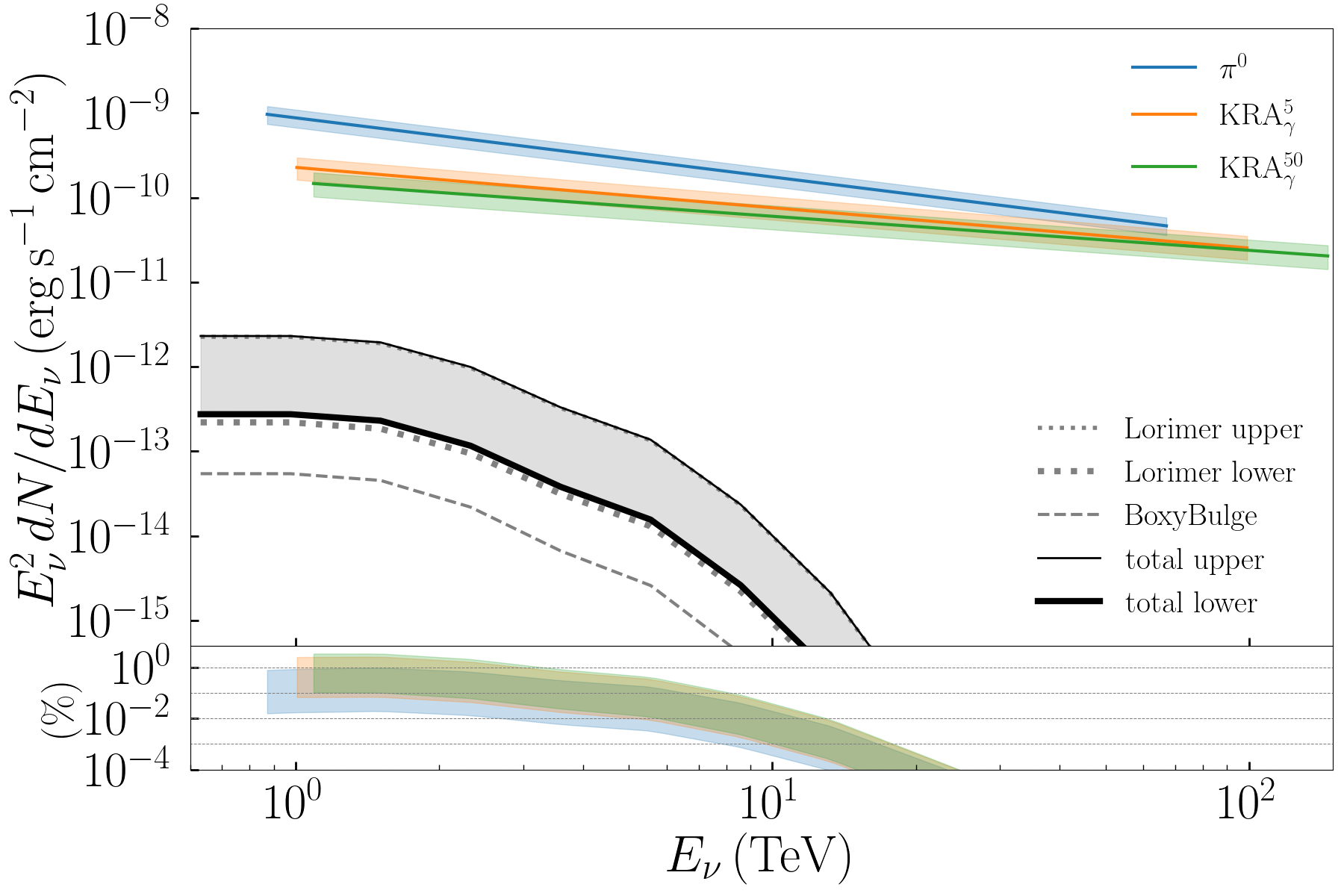}
    \caption{Predicted neutrino flux from the source populations as shown in the legend (also see Fig.~\ref{fig: spectral contribution of qbhxrbs Nustar}), assuming that the \blj jet scenario explains their jet dynamics. The subplot shows the contribution of the total predicted flux to the detected Galactic neutrino diffuse emission by IceCube \citep{icecube2023observation}. The three different colours represent the three different models used to extract the observed flux.}
    \label{fig: total neutrino flux}
\end{figure}

The recent Galactic neutrino diffuse emission detected by IceCube strengthens the idea that non-thermal accelerators exist in the Galactic plane \citep{icecube2023observation}. The hadronic processes that can take place in the jets of \bhs lead to the formation of non-thermal astrophysical neutrinos that reach energies of about a few dozen TeV. To determine the possible contribution of the q\bhs to the detected Galactic neutrinos, we self-consistently estimated the neutrino emission from each individual source for all the three source populations. In Fig.~\ref{fig: total neutrino flux}, we plot the total neutrino flux expected from the q\bhs in the Milky Way, and we compare it to the detected Galactic diffuse emission.
More precisely, the detected neutrino spectrum depends on the model, and for three different CR propagation models, \cite{icecube2023observation} derived three different neutrino spectra. The $\pi^0$ corresponds to the CR propagation model of \texttt{Galprop}, which uses as a normalisation the \fermi observations of the diffuse \gr emission \citep{Ackermann_2012}, whereas the KRA$_{\gamma}$ model corresponds to the CR propagation, for which the diffusion coefficient is spatially dependent, and there is an exponential cutoff in the propagating CR power-law at 5 or 50 PeV (KRA$_{\gamma}^5$ and KRA$_{\gamma}^{50}$, respectively; \citealt{gaggero2015KRAgamma}). 

We only show the case of \blj because neutrino emission in the \mlj scenario is further suppressed (see the discussion above for \grs). The total contribution of q\bhs to the detected spectrum is smaller than 1\% at an energy of about 1-2\,TeV, which rapidly drops to even less than $10^{-2}$\% at an energy of about 10\,TeV.

\section{Summary and conclusions}\label{Sec: summary and conclusions}

\bhs have recently been suggested as Galactic CR sources \citep{romero2003hadronic,romero2005misaligned,Torres_2005,BEDNAREK2005galactic,reynoso2008ss433,cooper2020xrbcrs,carulli2021neutrino,kantzas2023possible}. Most of these objects spend their lifetimes in quiescence, and only a handful of these are detected from the radio to X-rays. Recent observations of q\bhs suggest the existence of relativistic jets that can accelerate particles to high energies. We examined this scenario based on the well-studied q\bh \src. We used in particular two different jet models, \blj and \mlj, and we found that both explain the radio-to-X-ray electromagnetic spectrum well. Non-thermal relativistic electrons may reach energies of some dozen TeV that lead to some strong synchrotron and IC emission, which can explain the X-ray spectrum. Regardless of the acceleration of protons to non-thermal energies, we find that no significant \gr radiation from this source is produced that could be detected by current or planned facilities. 

The exact number of \bhs and q\bhs is debated. Observations and simulations yield values that range between a few thousand to up to $10^8$. Following the population synthesis analysis of \cite{Olejak2019synthesis}, we examined the contribution of $1.2\times 10^4-1.2\times 10^5$ q\bhs to the CR spectrum and the X-ray/\gr Galactic diffuse emission. While no significant contribution is expected in either the electron or proton CR spectra detected on Earth, we find that the cumulative intrinsic emission of these objects can contribute significantly to the Galactic diffuse emission. 
More precisely, assuming that all the q\bhs have the same spectral behaviour as \src but rescaled for the distance and viewing angle, we derived the predicted spectrum from X-rays to \grs. For a population of $10^4$ sources located in the boxy bulge and between $1.2\times 10^4-1.2\times 10^5$ in the Galactic disc, q\bhs can explain up to 10\% of the keV X-ray spectrum, depending on the jet conditions. This is the highest possible contribution, and more conservative numbers of q\bhs can reduce the contribution to a few percent. 
The soft MeV Galactic diffuse emission is expected to be dominated by cataclysmic variables and IC of CR electrons, but q\bhs can contribute up to 10-90 percent, depending on the jet dynamics and the exact number of sources. We find, moreover, that the contribution of q\bhs to the unresolved hard MeV spectrum decreases with energy because the emission originates in the non-thermal synchrotron emission of radiatively cooled electrons. This feature may help us to identify the origin of the MeV diffuse emission better. 
The radio counterpart of q\bhs may reveal their exact number. Current radio facilities that are sensitive to $\sim 10\,\rm \mu Jy$, such as MeerKAT, could detect some dozen q\bhs that can constrain their absolute number in the Milky Way better. In the more energetic regime, CTAO will detect some of these sources, in particular, those that reside at distances shorter than 2\,kpc and with viewing angles smaller than $15^{\circ}$ to allow for further boosting. These sources have only recently become detectable at distances as close as 0.5\, kpc and with viewing angles as low as $\sim 10^{\circ}$, and further observations would therefore help us to better constrain the contribution of \bhs and q\bhs in particular to the CR spectrum and the Galactic diffuse emission. Finally, q\bhs may explain up to 1\% of the diffuse neutrino background as detected by IceCube. Their contribution decreases significantly at neutrino energies beyond $\sim 10\,$TeV.

\begin{acknowledgements}

We are grateful for the very fruitful comments of the anonymous reviewer. DK and FC acknowledge funding from the French Programme d’investissements d’avenir through the Enigmass Labex, and from the ‘Agence Nationale de la Recherche’, grant number ANR-19-CE310005-01 (PI: F. Calore). We thank Thomas Siegert for useful discussions on the \spi analysis, and Marco Chianese on the non-thermal processes.

\end{acknowledgements}

\bibliography{references} 
\bibliographystyle{aa} 

\begin{appendix}
    
\section{Fermi spectra}\label{app: fermi diffuse emission for intermediate, ridge and off plane}

In Fig.~\ref{fig: spectral contribution of qbhxrbs Fermi intermediate, ridge and off plane}, we plot the contribution of the q\bhs to the \gr spectrum and compare it to the \fermi diffuse emission. See Fig.~\ref{fig: spectral contribution of qbhxrbs Fermi center} for details.

\begin{figure*}
    \centering

	\subfigure[Intermediate Galactic coordinates and \blj]{
        \includegraphics[width=0.975\columnwidth]{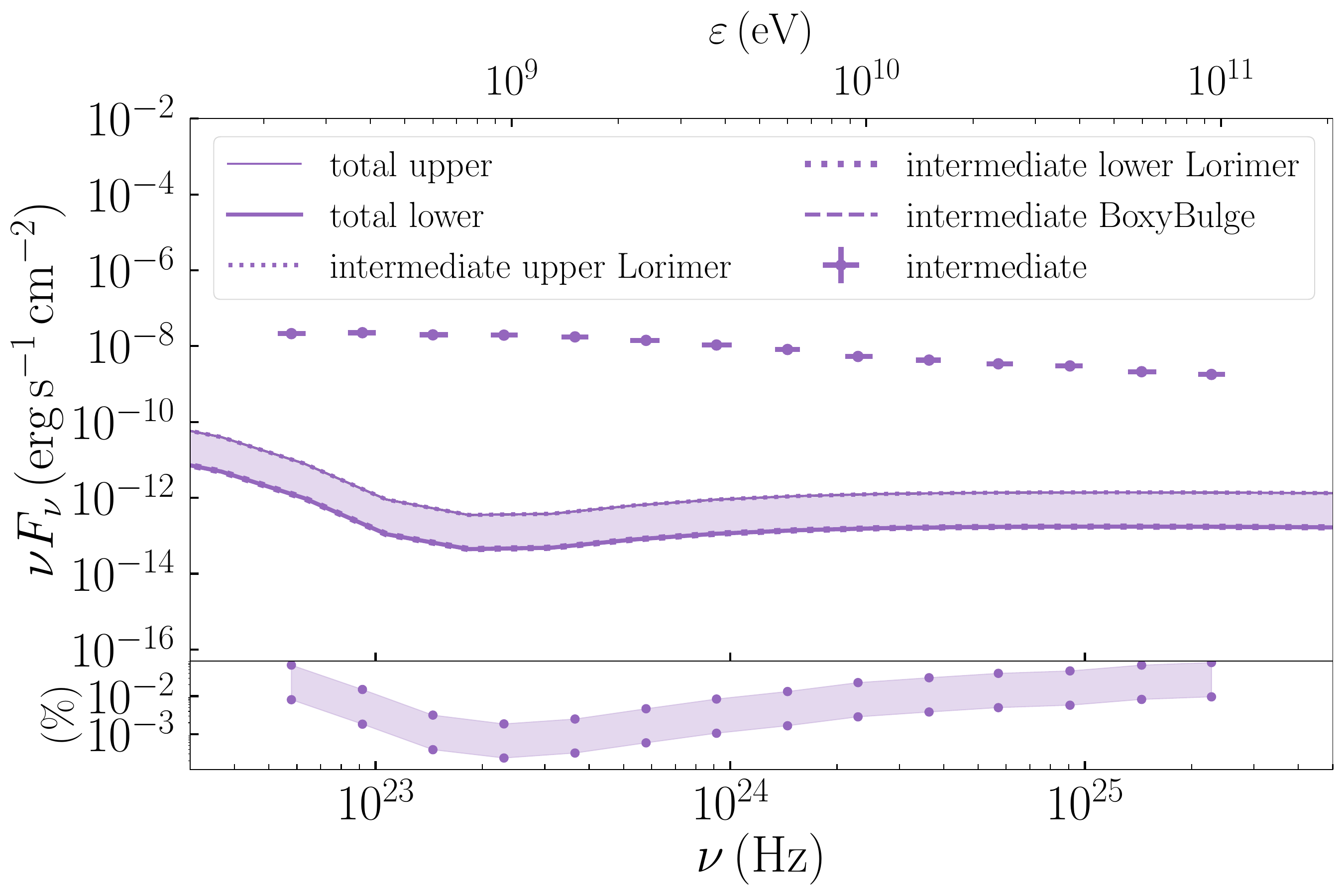}
    }
	\subfigure[Intermediate Galactic coordinates and \mlj]{
        \includegraphics[width=0.975\columnwidth]{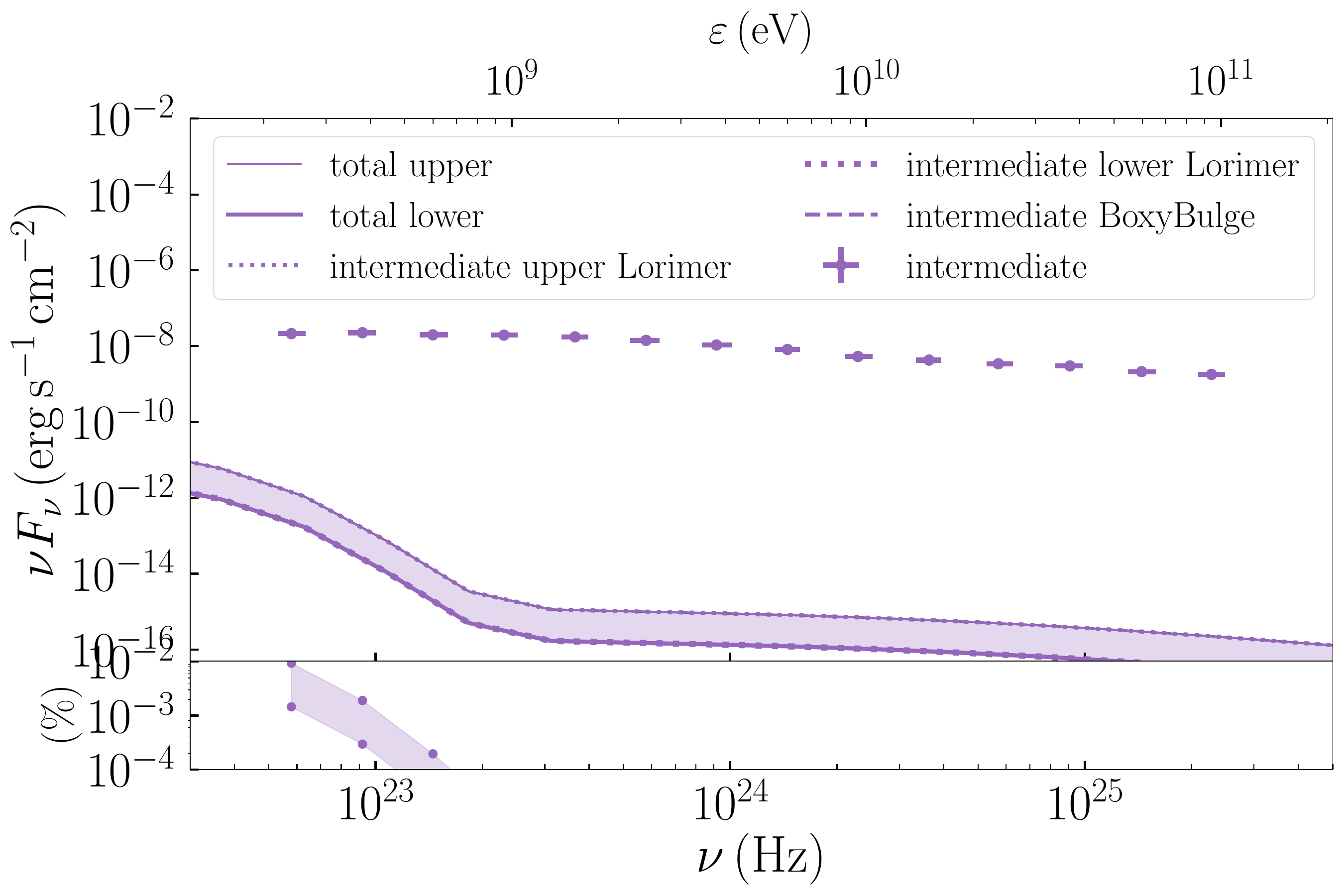}
    }
	\subfigure[The Galactic ridge and \blj]{
        \includegraphics[width=0.975\columnwidth]{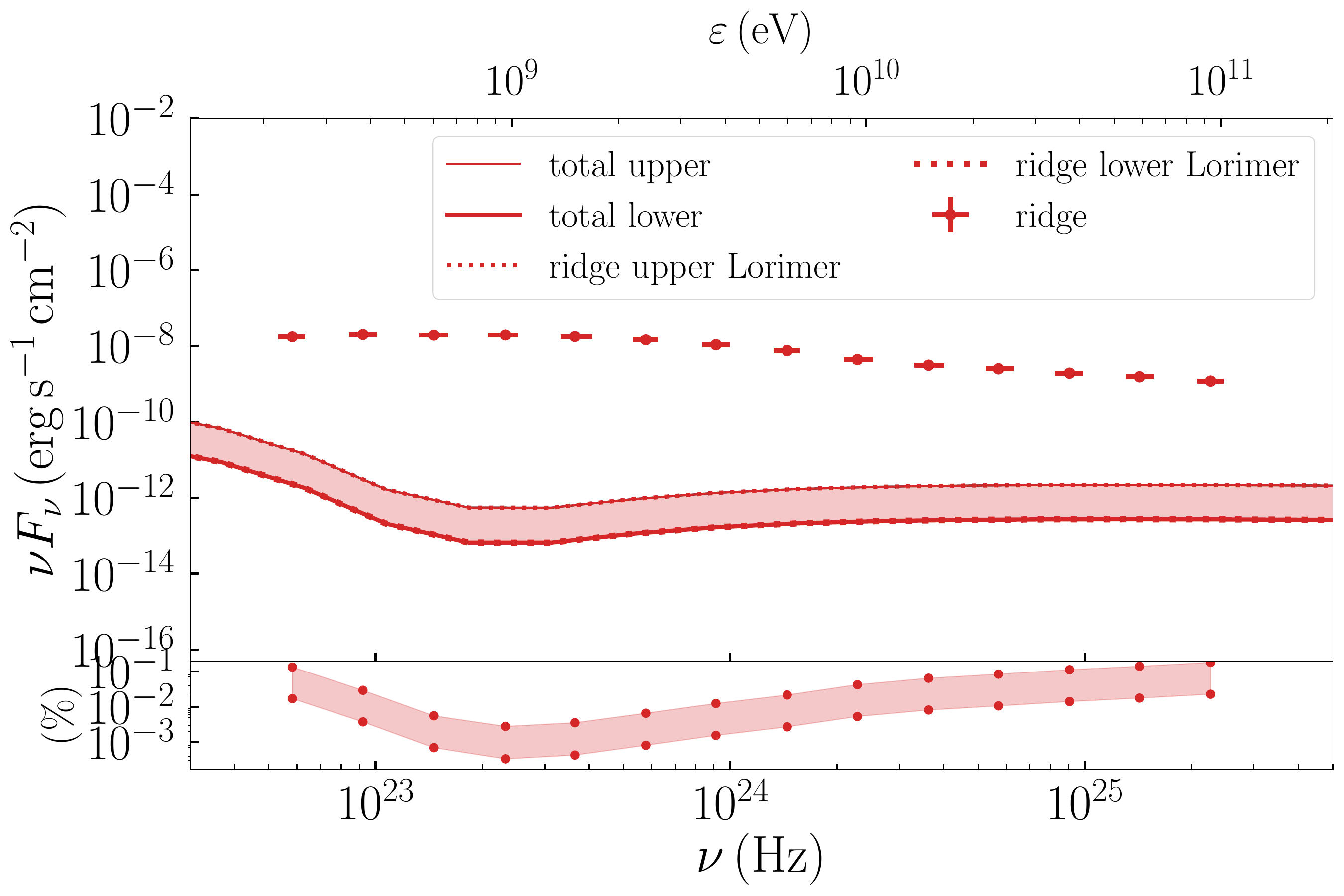}
    }
	\subfigure[The Galactic ridge and \mlj]{
        \includegraphics[width=0.975\columnwidth]{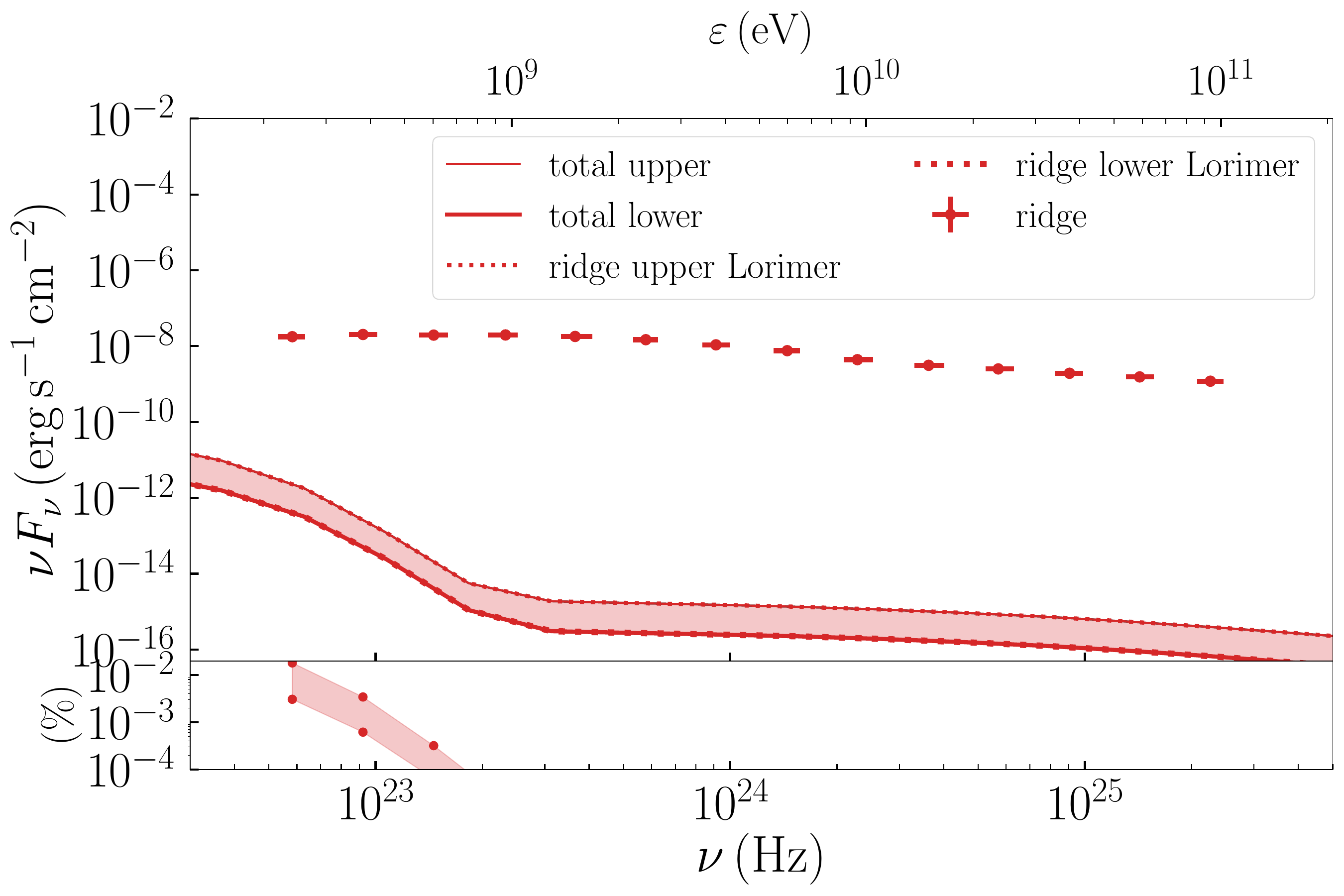}
    }

	\subfigure[Off the Galactic plane and \blj]{
        \includegraphics[width=0.975\columnwidth]{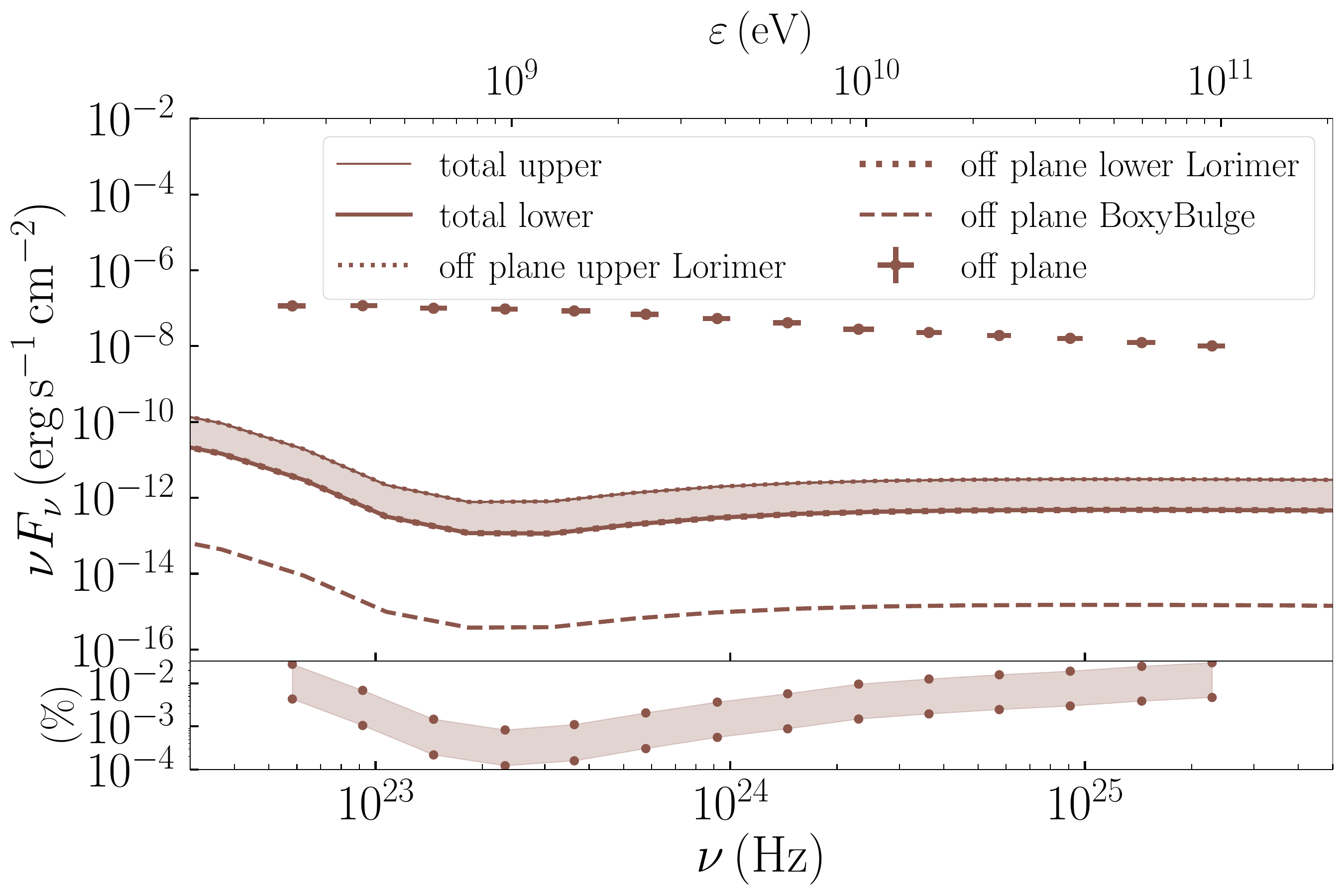}
    }
        \subfigure[Off the Galactic plane and \mlj]{
        \includegraphics[width=0.975\columnwidth]{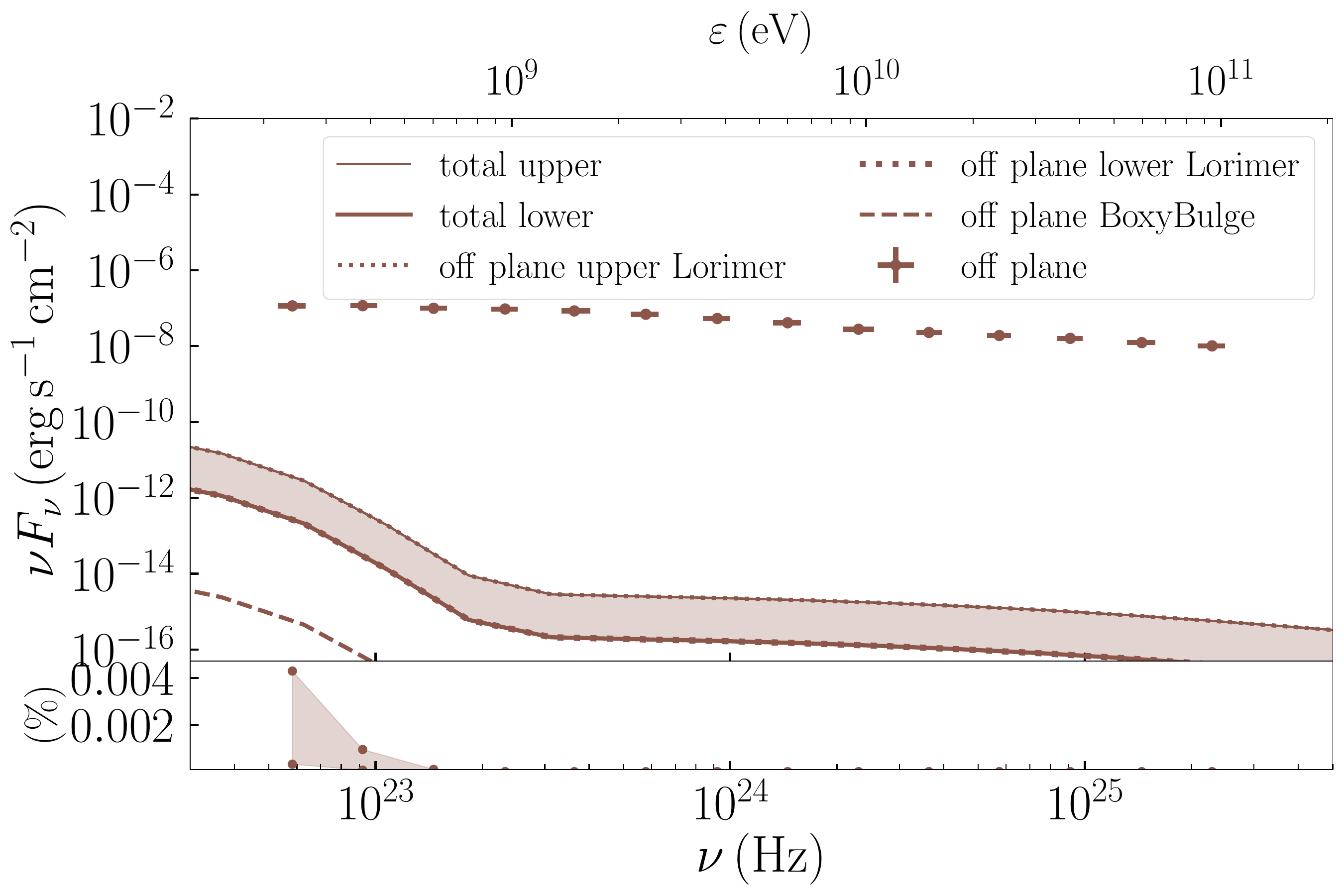}
    }
    \caption{ Similar to Fig.~\ref{fig: spectral contribution of qbhxrbs Fermi center} but for the \fermi detection of intermediate Galactic coordinates $10^\circ \leq |b|\leq 20^\circ$ in the upper panels, for the Galactic ridge with $80^\circ \leq |l| \leq 180^\circ$ and $|b|\leq 8^\circ$ in the middle panels, and off plane $ |b|\geq 8^\circ $ in the lower panels.
}
    \label{fig: spectral contribution of qbhxrbs Fermi intermediate, ridge and off plane}
\end{figure*}

\section{\dragon population setup} \label{app: the setup for dragon and the distributiions}

The total number of sources is the sum of the sources in the disc $N_d$, the sources located in the central parsec region $N_c$ add those in the bulge $N_b$:
\begin{equation}
    N_{\rm total} = N_d + N_c + N_b.
\end{equation}

Each population follows a distribution:
\begin{equation}
    N = \int_0^{\infty}N_0 \rho(R)\xi(z)dV,    
\end{equation}
where $N_0$ is the normalisation of each population, and $\rho(R)$ and $\xi(z)$ are the (unitless) number densities in the radial direction and perpendicular to the Galactic plane, respectively.

The disc sources follow the profile of the Lorimer distribution \citep[][]{Lorimer2006}:
\begin{equation}
    \rho_d(R)\xi_d(z) = \left(\frac{R}{R_0}\right)^B e^{-C(R-R_0)/R_0}\, e^{-|z|/\varepsilon},
\end{equation}
where $R_0=8.3$kpc, $B=1.9$, $C=5$, and $\varepsilon=0.18$kpc.
The normalisation of the Lorimer distribution is hence given by:
\begin{equation}
\begin{split}
    N_{0,d} & = \frac{N_d}{\dfrac{2\pi R_0^2 \varepsilon }{C^{B+2}e^{-C}} \bigintss _{\ 0}^{\infty} t^{B+1}e^{-t}dt, }\\
            & = \frac{N_d C^{B+2}e^{-C}}{2\pi R_0^2\varepsilon\ \Gamma(B+2)Q(B+2,0)}.
\end{split}
\end{equation}
where we set $t=cR/R_0$, $\Gamma$ is the Euler (gamma) function and $Q$ is the generalised incomplete gamma function. 

The sources of the Galactic centre follow a Gaussian distribution \citep[][]{Mori_2021} whose the normalisation is:
\begin{equation}
    N_{0,c} = \frac{N_c}{ \sqrt{2\pi} \varepsilon \left[ 2\sigma_c e^{-\mu_c^2/4\sigma_c^2} + \sqrt{\pi }\mu_c \left( \textrm{erf} \left( \dfrac{\mu_c}{2\sigma_c } \right) +1 \right)   \right] },
\end{equation}
where $\mu_c = 2\,$pc, $\sigma_c = 20\,$pc and $\textrm{erf}$ is the error function. This normalisation yields $N_{0,c} = 98040\times \dfrac{N_c}{1000}$.

Finally, the sources of the bulge follow a spherical distribution \citep[][]{korol2018multimessenger}:
\begin{equation}
    \rho_b(r) = e^{-r^2/2r_b^2},
\end{equation}
where $r_b = 0.5\,$kpc is the characteristic radius of the bulge and $r = \sqrt{R^2 +z^2}$ is the spherical distance from the Galactic centre. The normalisation of this distribution is:
\begin{equation}
    N_{0,b} = \frac{N_b}{4\pi \left(\sqrt{2}r_b\right)^3 \bigintss_{ \ 0}^{\infty} x^2 e^{-x} dx} = \frac{N_b}{8\pi \left(\sqrt{2}r_b\right)^3}.
\end{equation}

In a Cartesian 3D coordinate system, and for a more realistic distribution, we use the boxy bulge \citep[][]{cao2013new} that is described by the modified Bessel function of the second kind $K_0(r_s)$ where 
\begin{equation}
    r_s = \left( \left(  \left(\frac{x}{x_0}\right)^2   + \left(\frac{y}{y_0}\right)^2 \right)^2  + \left(\frac{z}{z_0}\right)^4\right)^{1/4},
\end{equation}
with $x_0 = 0.69\,$kpc, $y_0 = 0.29\,$kpc and $z_0 = 0.27\,$kpc. The normalisation for the 3D boxy bulge can be written as $N_{0,BB} = 7420\times \frac{N_{BB}}{10000}$. For all the normalisations we assume the Milky Way to extend to some radius of 12\,kpc and height 4\,kpc.

We define the fraction of sources that are located in the GC with respect to the total number of sources:
\begin{equation}
    \varepsilon = \frac{N_c}{N_{\rm total}},
\end{equation}
which we allow to vary.

The total source luminosity is 
\begin{equation}
    \begin{split}
        L_{\rm CR} & = \int N_0\rho(R)\xi(z)dV \int G(p) Q_0(p) f(E,p)dEdp,\\
        &= N_{\rm total}  \int \phi(f_{\rm esc}) f_{\rm esc}\ P_{\rm p}df_{\rm esc},
    \end{split}
\end{equation}
where $f_{\rm esc}$ is the fraction of CRs escaped from the jets of each individual source, $\phi(f_{\rm esc})$ is the PDF of $f_{\rm esc}$ and $P_{\rm p}$ is the power carried by the protons, which according to \src is $8.9\times 10^{34}\,\rm erg\,s^{-1}$. We calculate $N_0$ from the total number of sources
\begin{equation}
    \begin{split}
        N_0 &= \frac{N_{\rm total}}{\bigintss \rho(R) \xi(z)2\pi RdRdz},\\
            &=  N_{\rm total} \left(  \dfrac{N_c}{N_{0,c}}+\dfrac{N_d}{N_{0,d}} +\dfrac{N_b}{N_{0,b}}  \right)^{-1},
    \end{split}
\end{equation}
and $f(E)$ is the energy distribution of the particles of each source, which we assume to be a power-law $E^{-p}$ with an exponential cutoff at some maximum energy $E_{\rm max}$. The power-law index is around 2, and we assume that its probability density follows a Gaussian $G(p)$ with $\mu_p = 2.0$ and $\sigma_p = 0.3$ (cite particle acceleration papers here). In \dragon we use as an input the quantity $N_0 Q_0$ (\textit{Q0\_custom}), so we calculate $Q_0$ as
\begin{equation}
    Q_0(p) = \dfrac{\bigintss \phi(f_{\rm esc})f_{\rm esc}\ P_p df_{\rm esc}}{\bigintss f(E,p) dE}.    
\end{equation}

For the case of a power law $f(E) = E^{-p}$ from $E_{p,\rm min} = 1\,\rm GeV$ to $E_{p, \rm max} = 36\,\rm TeV$ (see Section where we describe the jet model and how we calculate the maximum energy for protons)

\section{CR spectrum}\label{app: CR proton and electron spectra}

\begin{figure}
    \centering
        \includegraphics[width=1.\columnwidth]{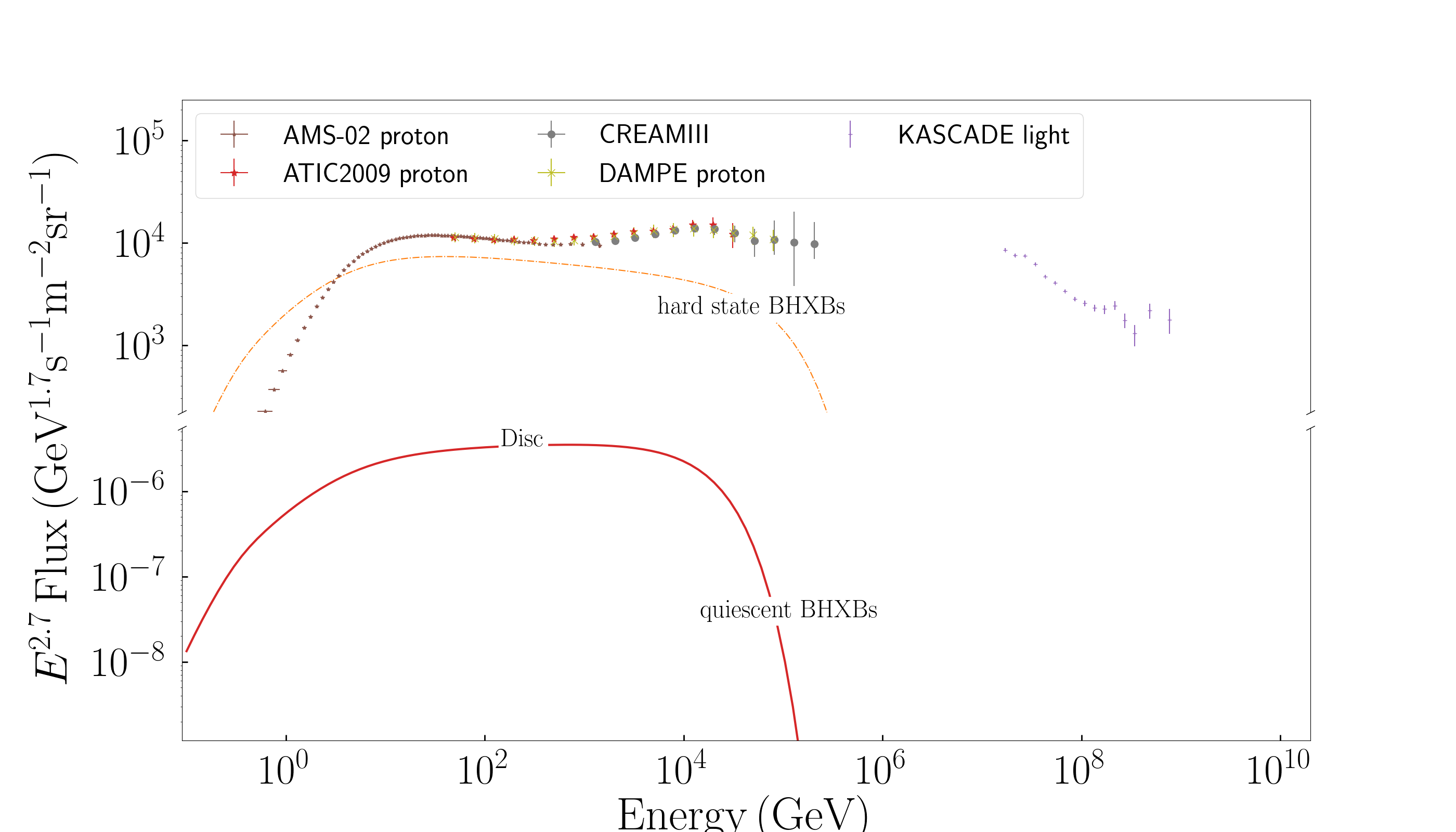}
    \caption{ The contribution of the q\bhs of a Lorimer-like distribution in the Galactic disc to the proton CR spectrum as detected on Earth. For comparison, we include the contribution of the \bhs in the hard state from \protect\cite{kantzas2023possible}.
    }
    \label{fig: CR p spectra}
\end{figure}

\begin{figure}
    \centering
      \includegraphics[width=1.\columnwidth]{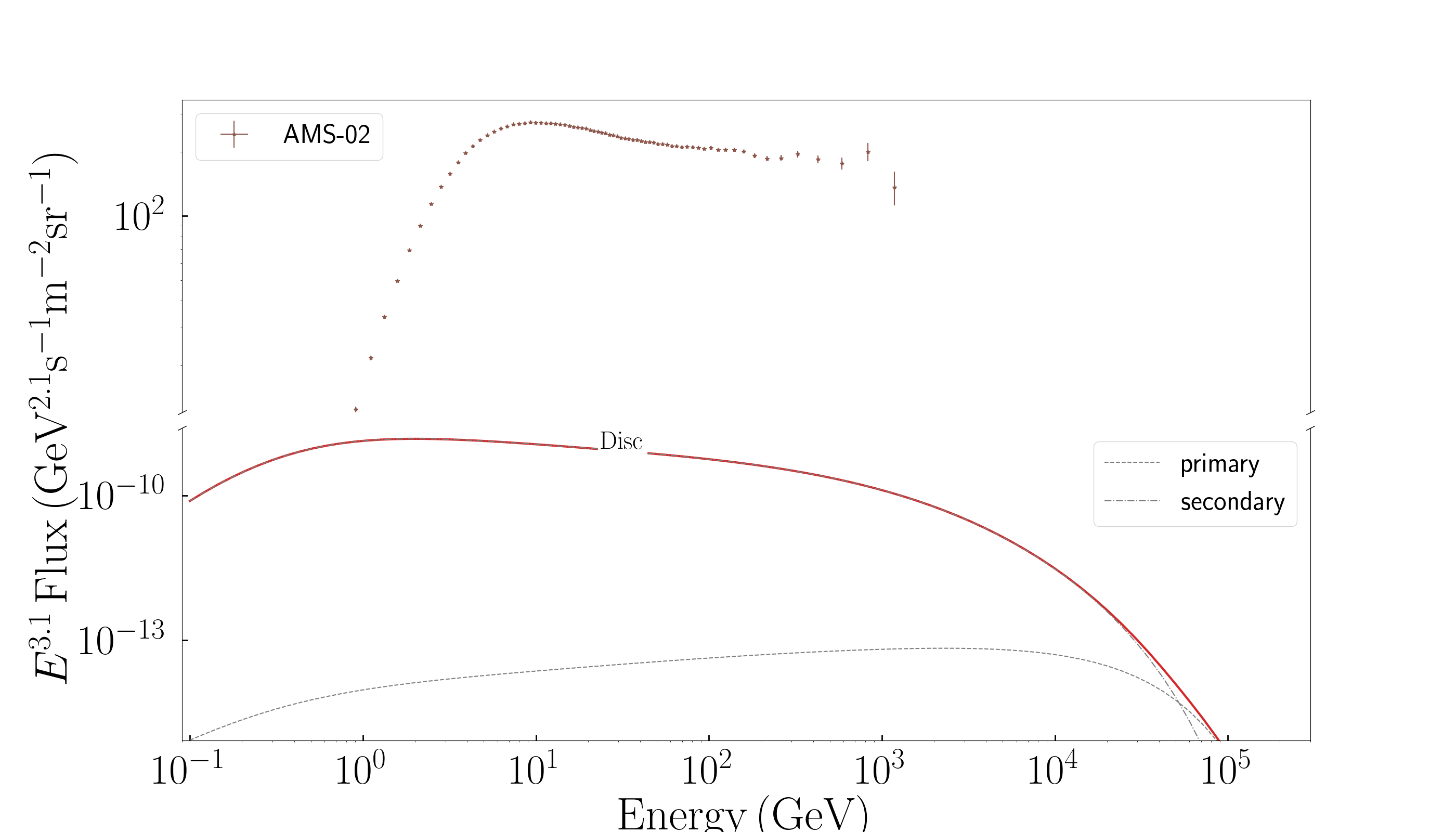}
    \caption{ The contribution of a Lorimer-like distribution of q\bhs in the Galactic disc to the electron CR spectrum detected on Earth. We plot both the primary electrons (dotted line), namely those that escape the q\bhs and propagate in the Galaxy, and the secondary electrons that are formed after the inelastic collisions of the propagating proton CRs with the Galactic gas (dotted line). 
    }
    \label{fig: CR e spectra}
\end{figure}

In Fig.~\ref{fig: CR p spectra}, we plot the proton CR spectrum where we show the contribution of the q\bhs to the detected spectrum. We compare to AMS-02 proton spectrum \citep{Aguilar2015amsH}, the ATIC proton data \citep{chang2008excess}, the CREAMIII \citep{Yoon_2017}, DAMPE \citep{dampe2019measurement}, and KASCADE \citep{Apel2011kascade}. The q\bhs mainly from the disc (because the boxy bulge GeV-TeV protons are subdominant) contribute very insignificantly, with less than $10^{-10}$. In Fig.~\ref{fig: CR e spectra} we show the electron CR spectrum using the AMS-02 electron data \citep{Aguilar2014AMSel}, and similar to protons, the contribution is negligible. In the same plot, we show the contribution of both the primary electrons and those produced by the propagation of protons and their interaction with the Galactic gas. We see that the secondary electrons dominate over the primary, but still are not enough to contribute somehow to the detected electron spectrum.

\end{appendix}

\end{document}